\documentclass[journal]{IEEEtran}
\IEEEoverridecommandlockouts

\usepackage[pdftex]{graphicx}
\usepackage{algorithm,algpseudocode}

 \usepackage{cite}
\usepackage[cmex10]{amsmath}
\usepackage{amssymb}  
\usepackage{amsxtra}
\usepackage{amscd}
\usepackage{amsthm}
 \usepackage{amsxtra}
 \usepackage{amscd}
 \usepackage{amsthm}
\usepackage{tikz}
\usetikzlibrary{matrix}
\usepackage{graphicx, subfigure}   
\usepackage{comment}
\usepackage[pdftex]{graphicx}
\usepackage{verbatim} 
\usepackage{array}  
\usepackage{authblk}
\usepackage{multirow}  
\usepackage{arydshln}
\usepackage{hhline}

\usepackage{color}

\def\BibTeX{{\rm B\kern-.05em{\sc i\kern-.025em b}\kern-.08em
    T\kern-.1667em\lower.7ex\hbox{E}\kern-.125emX}}

\usepackage{color}

\renewcommand{\thefootnote}{\arabic{footnote}}
\usepackage{balance}
\usepackage[keeplastbox]{flushend}

  \author{Ali Tugberk Dogukan,~\IEEEmembership{Student Member,~IEEE} and Ertugrul Basar,~\IEEEmembership{Senior Member,~IEEE}  
  	\vspace{-0.8cm}
\thanks{Manuscript received December 20, 2019; revised March 20, 2020; revised    May 28, 2020; accepted July 17, 2020. Date of publication XXXX XX, 2020; date of current version XXXX XX, 2020. This work was supported by the Scientific and Technological Research Council of Turkey (T\"{U}B\.{I}TAK) under Grant 218E035. The associate editor coordinating the review of this paper and approving it for publication was Prof. Zhaoyang Zhang.
	
The authors are with the Communications Research and Innovation Laboratory (CoreLab),  Department of Electrical and Electronics Engineering, Ko\c{c} University, Sariyer 34450, Istanbul, Turkey. \mbox{Email: adogukan18@ku.edu.tr, ebasar@ku.edu.tr}

Color versions of one or more of the figures in this paper are available online at http://ieeexplore.ieee.org.

Digital Object Identifier 10.1109/TWC.2020.3010839

} 
}

\begin{document}
	
	\title{{ Super-Mode OFDM With Index Modulation }}
	\maketitle	
	
	\begin{abstract}

Orthogonal frequency division multiplexing (OFDM) with index modulation (OFDM-IM) appears as a promising multi-carrier waveform candidate for beyond 5G due to its attractive advantages such as operational flexibility and ease of implementation. However, OFDM-IM may not be a proper choice for 5G services such as enhanced mobile broadband (eMBB) since achieving high data rates is challenging because of its null subcarriers. One solution to enhance the spectral efficiency of OFDM-IM is the employment of multiple distinguishable constellations (modes) by also exploiting its null subcarriers for data transmission. This paper proposes a novel IM technique called super-mode OFDM-IM (SuM-OFDM-IM), where mode activation patterns (MAPs) and subcarrier activation patterns (SAPs) are jointly selected and conventional data symbols are repetition coded over multiple subcarriers to achieve a diversity gain. For the proposed scheme, a low-complexity detector is designed, theoretical analyses are performed and a bit error rate (BER) upper bound is derived. The performance of the proposed system is also investigated through real-time experiments using a software-defined radio (SDR) based prototype. We show that SuM-OFDM-IM exhibits promising results in terms of spectral efficiency and error performance; thus, appears as a potential candidate for 5G and beyond communication systems.

	\end{abstract}

	\IEEEpeerreviewmaketitle
		\IEEEpubidadjcol
		
	\renewcommand{\thefootnote}{\fnsymbol{footnote}}

	\section{Introduction}
	\IEEEPARstart{I}{ndex} modulation (IM)\cite{IM}, which employs the indices of transmit entities such as antennas, time slots and radio frequency (RF) mirrors,  has attracted significant attention from the researchers due to its advantages over conventional communication systems in terms of transceiver complexity and spectral/energy efficiency. These attractive features of IM make it a potential candidate for  communication systems beyond 5G. IM has been employed in numerous dimensions as mentioned above, and it can also be implemented in the frequency domain by utilizing the indices of the  available subcarriers of multi-carrier systems\cite{basar2017index, mao2018novel}.
	
	In \cite{OFDMIM}, orthogonal frequency division multiplexing (OFDM) has been effectively combined with IM by activating a number of subcarriers in a subblock while de-activating the remaining ones. In OFDM-IM, information bits are conveyed by not only $Q$-ary modulated symbols but also active subcarrier indices. Compared to conventional OFDM, it is shown that OFDM-IM is capable of providing a better error performance while consuming less power due to its null subcarriers. The error performance of OFDM-IM can be improved by employing a subcarrier-level block interleaver \cite{INTERLEAVE}. To further enhance the error performance of OFDM-IM, coordinate interleaved OFDM-IM (CI-OFDM-IM), which provides a diversity gain by transmitting the real and imaginary parts of modulated symbols over different active subcarriers, is proposed in \cite{CIOFDMIM}.  In \cite{OFDMSNM}, OFDM with subcarrier number modulation (OFDM-SNM) is proposed where the number of active subcarriers in a subblock is determined by the incoming bits. A clever technique that decreases the modulation order while increasing the number of bits transmitted by IM without any reduction in spectral efficiency, is proposed \cite{LOWER}. An energy- and spectrum-efficient system that employs unused subcarrier activation patterns (SAPs) in OFDM-IM by selecting a certain number of adjacent subblocks jointly, is proposed in \cite{JOINT}. In recent years, OFDM-IM has also been applied to emerging communication systems such as dual-hop networks \cite{dang2017adaptive,dang2018outage}, multiple-input multiple-output (MIMO) systems \cite{basar2016multiple,zheng2017multiple,ozturk2019multiple,lu2018compressed}, cognitive radio \cite{ma2017ofdm,li2019opportunistic} and optical communications \cite{bacsar2015optical}.
	 
	\begin{figure*}[t]
		\centering
		\includegraphics[width=0.58\linewidth ]{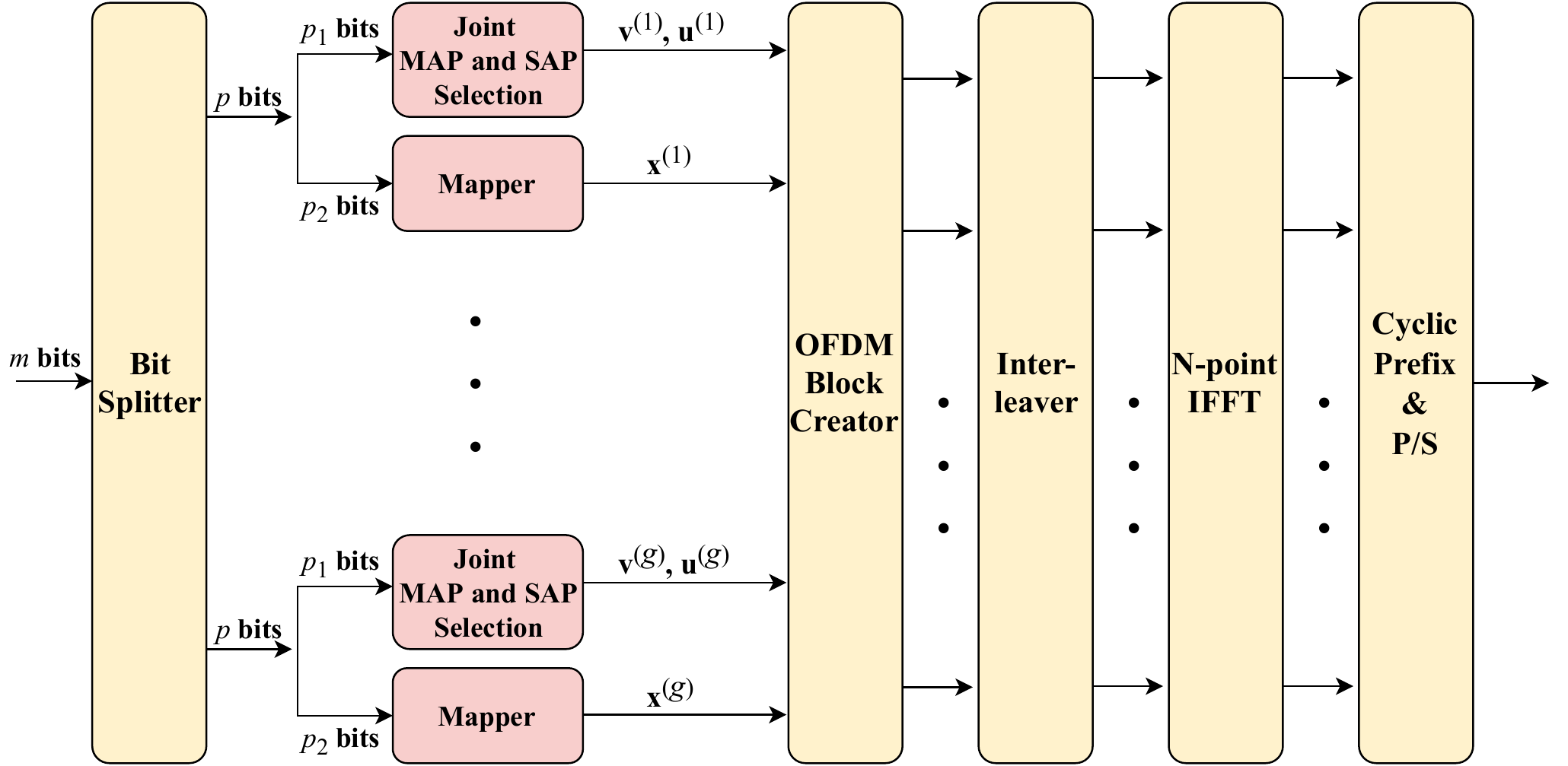}	
		\caption{Transmitter Structure of SuM-OFDM-IM.}
		\vspace{-0.4cm}
		\label{Fig. 1}
	\end{figure*}

	Because of its null subcarriers, achieving very high data rates with OFDM-IM is not possible and this may be critical for future enhanced mobile broadband (eMBB) applications such as augmented/virtual reality (AR/VR) and $4$K/$8$K video streaming. To overcome this critical drawback, multiple constellations are utilized by activating not only the selected subcarriers but also the null ones. Dual-mode aided OFDM (DM-OFDM), which also utilizes the null subcarriers to transmit data symbols drawn from a secondary distinguishable constellation, is proposed in \cite{DUAL}. In \cite{TDOFDM}, a diversity gain is achieved by transmitting the same data symbols through re-activation of the null subcarriers with a second mode. DM-OFDM is generalized in \cite{GDUAL} by changing the number of subcarriers drawn from the same constellation. The performance of DM-OFDM and OFDM-IM has been investigated in real-time by adopting software defined radios (SDRs) in \cite{PRACT}. In DM-OFDM, only two different constellations are considered; however, multiple-mode OFDM-IM (MM-OFDM-IM), which exploits all $n$ subcarriers in each subblock with $n$ different distinguishable modes, is proposed to employ more number of constellations \cite{MMOFDMIM}. MM-OFDM-IM uses the full permutations of these constellations to apply IM. In \cite{GMM}, a generalized version of MM-OFDM-IM is proposed where the size of the constellations can be altered although there is no increase in the number of bits conveyed by IM. To obtain diversity gains, coordinate interleaved MM-OFDM-IM (CI-MM-OFDM-IM) and linear constellation precoded MM-OFDM-IM (LCP-MM-OFDM-IM) schemes have been presented \cite{DEMM}. Although these noticeable improvements, there is still an undeniable need for multi-carrier waveforms that provide both reliable error performance and high data rate for future eMBB services. For this reason, our motivation is to exploit IM to further increase the spectral efficiency of OFDM-based waveforms while providing a high reliability.
	
	In this paper, we propose a novel OFDM-based transmission system called \textit{super-mode OFDM-IM (SuM-OFDM-IM)}, which is capable of yielding both a diversity gain and a high spectral efficiency. In SuM-OFDM-IM, according to the incoming bits, mode activation patterns (MAPs) and SAPs are jointly determined; therefore, additional information bits can be transmitted by IM. { We also consider determining MAPs and SAPs separately to provide an alternative solution and call this scheme  \textit{separated SuM-OFDM-IM (S-SuM-OFDM-IM)}.} It is well known that the bits transmitted by IM are more reliable than the ones transmitted by conventional $Q$-ary PSK/QAM. In other words, conventional PSK/QAM symbols restrict the diversity gain of the system. In our scheme, we apply repetition coding to the modulated symbols drawn from the selected modes over multiple subcarriers to attain a diversity gain. We propose a novel encoding technique as well as a log-likelihood ratio (LLR)-based reduced complexity maximum likelihood (ML) detection algorithm. Additionally, real-time error performance of the proposed system is investigated through a practical testbed using SDR units. 
	
	The rest of the paper is organized as follows. In Section II, we present the system model of SuM-OFDM-IM. Section III deals with the selection of modes. In Section IV, we derive an upper bound on the bit error rate (BER). Monte Carlo simulation results and the complexity analysis are given in Section V. The practical implementation of the proposed system is shown in Section VI and finally, we conclude the paper in Section VII.

\section{System Model}
In this section, we first introduce an encoding algorithm for SuM-OFDM-IM, which jointly determines MAP and SAP in order to increase the number of bits conveyed by IM. We also consider a clever symbol assignment strategy to provide a diversity gain. Second, we propose a novel LLR-based reduced complexity ML detector. 
\vspace{-0.3cm}
\subsection{Transmitter}
SuM-OFDM-IM is an OFDM-based transmission system with $N$ subcarriers whose transmitter structure is given in Fig. 1. A total of $m$ bits are conveyed for each symbol transmission and split into groups with $p=m/g$ bits, where $g$ is the number of subblocks. Since the same procedures are applied for each subblock, for convenience, we consider the mapping of $p$ bits into the $\alpha$th subblock $\mathbf{s}^{(\alpha)}$, where $\alpha\in\{1,\dots,g\}$. These $p$ bits are split into two parts:
\begin{enumerate}
    \item $p_1=\lfloor{\log_2(\binom{M}{2}\binom{n}{n/2})}\rfloor$
    \item $p_2=\frac{n}{2}\log_2Q$
\end{enumerate}
where $n$, $M$ and $Q$ are the size of the subblock, $n=2^r, r\in\{2,3,\dots\}$, the total number of modes, and constellation size, respectively, and $\binom{\cdot}{\cdot}$ is the binomial coefficient. First, $p_1$ bits jointly determine an MAP $(\mathbf{v}^{(\alpha)}=[v_1, v_2], v_l\in\{1,\dots,M\}, l=1,2)$ and an SAP $(\mathbf{u}^{(\alpha)}=[u_1,\dots,u_{n/2}], u_m\in\{1,\dots,n\}, m=1,\dots,n/2)$, where we discuss this joint selection in the sequel. Here, the selected MAP includes two different modes to be able to employ IM $(v_1\neq v_2)$ and the strategy for mode selection is discussed in Section IV. We also activate $n/2$ subcarriers per mode to maximize $p_1$. The remaining indices that are not included in a selected SAP are represented by $\mathbf{w}^{(\alpha)}=[u_{n/2+1},\dots,u_n]$. The subcarrier indices $\mathbf{u}^{(\alpha)}$ and $\mathbf{w}^{(\alpha)}$ are exploited for the symbols drawn from the first and second modes, respectively. Then, since each symbol is repetition coded over a pair of subcarriers, $\mathbf{u}^{(\alpha)}$ and $\mathbf{w}^{(\alpha)}$ are split into $n/4$ groups $(\mathbf{u}_k^{(\alpha)}=[u_{2k-1},u_{2k}], \mathbf{w}_k^{(\alpha)}=[u_{n/2+2k-1},u_{n/2+2k}], k=1,\dots,n/4$. Second, $\frac{p_2}{2}=\frac{n}{4}\log_2Q$ bits determine the first $n/4$ modulated symbols $(\mathbf{x}_1^{(\alpha)}=[x_1,\dots,x_{n/4}])$, which are drawn from $\mathcal{M}_{v_1}$, $\mathcal{M}_{i}=\{\chi_{i1}, \dots, \chi_{iQ}\}, i\in\{1,\dots,M\}$, where $\chi_{iq}$ is the $q$th symbol of the $i$th mode and the set of all modes is $\mathcal{M}=\{\mathcal{M}_1, \dots,\mathcal{M}_M\}$. The average symbol power of $\mathcal{M}$ is normalized to unity. Finally, $\frac{p_2}{2}=\frac{n}{4}\log_2Q$ bits determine the last $n/4$ modulated symbols $(\mathbf{x}_2^{(\alpha)}=[x_{n/4+1},\dots,x_{n/2}])$, which are drawn from $\mathcal{M}_{v_2}$, and we obtain $\mathbf{x}^{(\alpha)}=[\mathbf{x}^{(\alpha)}_1 \mathbf{x}^{(\alpha)}_2]$. The $k$th element of $\mathbf{x}_1^{(\alpha)}$ and $\mathbf{x}_2^{(\alpha)}$ are repetition coded over the subcarriers of the target subblock with the indices given by $\mathbf{u}_k^{(\alpha)}$ and $\mathbf{w}_k^{(\alpha)}$, respectively. By performing the same steps for all elements of $\mathbf{x}_1^{(\alpha)}$ and $\mathbf{x}_2^{(\alpha)}$, $\mathbf{s}^{(\alpha)}$ is created. Consequently, a total of 
\begin{equation}
    p=\Bigl\lfloor{\log_2\Bigl(\binom{M}{2}\binom{n}{n/2}\Bigl)}\Bigr\rfloor+\frac{n}{2}\log_2Q
    \label{Eq1}
\end{equation}
bits are transmitted per subblock. 

The aim of this joint MAP and SAP selection is to further increase the number of bits conveyed by IM. Our joint selection algorithm of MAP and SAP is summarized below:
\begin{enumerate}
    \item Convert $p_1$ bits to a decimal number: $d$.
    \item Find the unique values $(a_1,a_2)$ satisfying the equation:\break 
    \begin{equation}
        d=a_1+\binom{M}{2}a_2,
        \label{Eq2}
    \end{equation}
    where $a_1\in\{0,\dots,\binom{M}{2}-1\}$ and $a_2\in\{0,\dots,\binom{n}{n/2}-1\}$ are the indices of the selected MAP and SAP, respectively.
    \item Determine MAP $(\mathbf{v}^{(\alpha)})$ and SAP $(\mathbf{u}^{(\alpha)})$ by applying $a_1$ and $a_2$ to the combinatorial algorithm \cite{OFDMIM}, respectively. 
\end{enumerate}
\begin{table}[t]
\renewcommand{\arraystretch}{1.1}%
      	\centering
      	\caption{Joint Selection of MAP and SAP}
      	\begin{tabular}{|>{\centering\arraybackslash}m{1cm} |>{\centering\arraybackslash}m{1cm}|>{\centering\arraybackslash}m{1cm} |>{\centering\arraybackslash}m{1cm}|>{\centering\arraybackslash}m{1cm}|} 
      	
      		\hline
      		$a_1$ & $[v_1,v_2]$ & $a_2$ & $[u_1,u_2]$\\ [0.5ex]
      		
      		\hline\hline
      		$0$ & $[1,2]$ & $0$ & $[1,2]$ \\\hline
        	$1$ & $[1,3]$ & $1$ & $[1,3]$ \\\hline
      	    $2$ & $[2,3]$ & $2$ & $[2,3]$ \\\hline
            $3$ & $[1,4]$ & $3$ & $[1,4]$ \\\hline
            $4$ & $[2,4]$ & $4$ & $[2,4]$ \\\hline
            $5$ & $[3,4]$ & $5$ & $[3,4]$ \\\hline

      	\end{tabular}
      	\label{ciz1}
      	\vspace{-0.4cm}
\end{table}

\textit{Example:} Assuming $M=4, n=4, Q=4$, the number of bits conveyed by IM is $p_1=\lfloor{\log_2(\binom{4}{2}\binom{4}{4/2})}\rfloor=5$ and the number of bits conveyed by conventional data symbols is $p_2=\frac{4}{2}\log_2(4)=4$. For this case,  a total of $\binom{4}{2}=6$ MAPs and $\binom{4}{2}=6$ SAPs are obtained as in Table I. There are $36$ possible pairs of $(a_1,a_2)$ and $32$ of these are employed to transmit $\lfloor{\log_2(36)}\rfloor=5$ bits. Let $p_1$ and $p_2$ be $\{01001\}$ and $\{1110\}$, respectively. The $(a_1,a_2)$ values satisfying (\ref{Eq2}) are $(3, 2)$. {In the absence of joint selection, only four bits can be transmitted per subblock instead of five by the indices of modes and subcarriers for these given parameters. Therefore, an additional bit can be conveyed by the joint selection in this specific case.} Then, MAP is determined as $\mathbf{v}=[1,4]$, i.e, $\mathcal{M}_1$ and $\mathcal{M}_4$ are employed. SAP is determined as $\mathbf{u}=[1,3]$, so the remaining indices are $\mathbf{w}=[2,4]$. For simplicity, we have removed the superscript $(\alpha)$. The first two bits of $p_2$, which are $\{11\}$, select the fourth symbol of the first selected mode $(\chi_{14})$ and the second two bits of $p_2$, which are $\{10\}$, select the second symbol of the second selected mode $(\chi_{42})$. Finally, $\chi_{14}$ and $\chi_{42}$ are placed into the indices of subblock entries with  $\mathbf{u}$ and $\mathbf{w}$, respectively and the overall subblock is obtained as:
\begin{equation}
	\mathbf{s}=[\chi_{14} \ \chi_{42} \ \chi_{14} \ \chi_{42}]^{\mathrm{T}}.
	\label{Eq3}
\end{equation}

Having constructed the OFDM symbol subblock by subblock, a block-type interleaver is executed as in \cite{MMOFDMIM} and the same procedures applied in OFDM are performed. The inverse fast fourier transform (IFFT) algorithm is applied to the OFDM block to obtain the time domain OFDM block as:
\begin{equation}
    \mathbf{s}_T=\mathrm{IFFT}\big\{\mathbf{s}_F\big\} = \begin{bmatrix} S(1) & S(2) & \cdots & S(N) \end{bmatrix}^\mathrm{T}.
    \label{Eq4}
\end{equation}
\noindent
Then, a cyclic prefix (CP) of length $L$ samples $\begin{bmatrix} S(N-L+1) & \cdots & S(N-1)S(N) \end{bmatrix}^\mathrm{T}$ is added to the beginning of the OFDM block. After parallel to serial (P/S) and digital/analog conversions are applied, the signal is transmitted over a frequency-selective Rayleigh fading channel with the channel impulse response (CIR) coefficients
\begin{equation}
\mathbf{c}_T=\begin{bmatrix} c_T(1) & \cdots & c_T(v) \end{bmatrix}^\mathrm{T},
\label{Eq5}
\end{equation}
where $c_T(\beta),\beta=1,\dots,v$ are circularly symmetric complex Gaussian random variables with the $\mathcal{C}\mathcal{N}(0,\frac{1}{v})$ distribution. With the assumption that the channel remains constant during transmission of an OFDM block and the CP length $L$ is larger than $v$, the equivalent frequency domain input-output relationship of this OFDM scheme is given by 
\begin{equation}
    y_F(\beta)=s_F(\beta)c_F(\beta)+w_F(\beta), \hspace{0.2cm}       \beta=1,\dots,N
    \label{Eq6}
\end{equation}
where $y_F(\beta)$, $c_F(\beta)$ and $w_F(\beta)$ are the received signals, the channel fading coefficients and the noise samples in the frequency domain, whose vector forms are given as $\mathbf{y}_F$, $\mathbf{c}_F$ and $\mathbf{w}_F$, respectively. The distributions of $c_F(\beta)$ and $w_F(\beta)$ are $\mathcal{C}\mathcal{N}(0,1)$ and $\mathcal{C}\mathcal{N}(0,N_{0})$, respectively, where $N_{0}$ is the noise variance in the frequency domain, which is equal to the noise variance in the time domain. The signal-to-noise ratio (SNR) is defined as $\rho=E_b/N_{0}$, where $E_b=(N+L)/ m$ is the average transmitted energy per bit. For simplicity, the effect of CP on the spectral efficiency is ignored. Then, the spectral efficiency of the proposed scheme is obtained as
\begin{equation}
    \eta =m/N \hspace{0.2cm} \text{[bits/s/Hz]}. 
    \label{Eq7}
\end{equation}

\subsubsection{Separate selection of MAPs and SAPs}
{Instead of joint selection, it is possible to determine MAPs and SAPs, separately. This scheme is called S-SuM-OFDM-IM. The only difference from SuM-OFDM-IM is the selection procedure of MAPs and SAPs. In this case, a total number of 
	\begin{equation}
		p=\Big\lfloor{\log_2\Big(\binom{M}{2}\Big)}\Big\rfloor+\Big\lfloor{\log_2\Big(\binom{n}{n/2}\Big)}\Big\rfloor+\frac{n}{2}\log_2Q
		\label{Eq8}
	\end{equation}
	bits are transmitted per subblock. As seen from Fig. 2, the joint selection provides higher spectral efficiency than the separate one for varying $Q$ values. However, in Section IV, we provide an interesting trade-off between error performance and spectral efficiency.}
\begin{figure}[!t]
	
		\centering
		\includegraphics[width=0.8\linewidth]{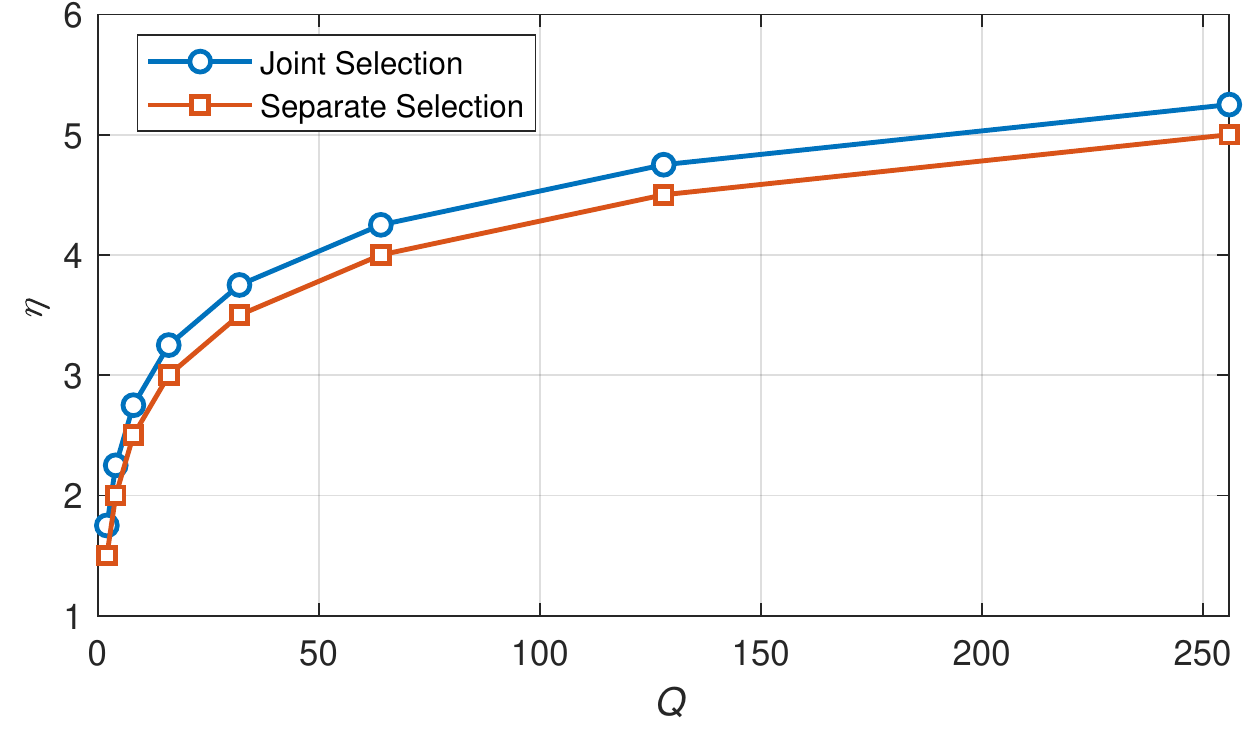}	
		\caption{Spectral efficiency comparison of joint and separate selections of MAPs and SAPs for varying values of $Q$ when $n=4$ and $M=4$.}
		\vspace{-0.4cm}
		\label{Fig. 1}
\end{figure}
 
\vspace{-0.2cm}
\subsection{Receiver} 

 \begin{algorithm}
	\normalsize 
	\caption{LLR-Based Reduced Complexity ML Detector $(n\in\{4,8\})$}
	\begin{algorithmic}[1]
		\renewcommand{\algorithmicrequire}{\textbf{Input:}}
		\renewcommand{\algorithmicensure}{\textbf{Output:}}
		\Require $\mathbf{y}_F^{(\alpha)}$, $\mathbf{c}_F^{(\alpha)}$, $N_0$, $\mathcal{M}$, $M$, $Q$, $p_1$, $n$
		\Ensure $\hat{a_1}, \hat{a_2}, \hat{\mathbf{x}}$
		\State $\mathbf{y}:=\mathbf{y}_F^{(\alpha)}, \mathbf{c}:=\mathbf{c}_F^{(\alpha)}$
		\For {$i = 1$ \textbf{to} $\binom{n}{n/2}$}
		\State $\mathcal{I}^{(i)}:=$combinatorics\_n2c($i-1$,$n/2$)
		\State $j:=1$
		\For {$m = 1$ \textbf{to} $M$}
		\For {$h = 1$ \textbf{to} $n/4$}
		\For {$q = 1$ \textbf{to} $Q$}
		\For {$z=1$ \textbf{to} $2$}
		\State $\hat{\mathcal{I}}:=\mathcal{I}^{(i,z+2(h-1))}+1$
		\State {\footnotesize $\delta_1^{(q,z+2(h-1))}:=-|\mathbf{y}^{(\hat{\mathcal{I}})}-\mathbf{c}^{(\hat{\mathcal{I}})}\mathcal{M}^{(m,q)}|^2/N_{0}$}
		\State
		{\footnotesize $\delta_2^{(q,h)}:=\delta_1^{(q,2h-1)}+\delta_1^{(q,2h)}$}
		\EndFor
		\EndFor
		\EndFor
		
		\For{$g_1=1$ \textbf{to} $n/4$}
		\For{$g_2=1$ \textbf{to} $g_1$}
		\If{$n=8$ $\&\&$ $g_1\neq g_2$}
		\For{$q_1=1$ \textbf{to} $Q$}
		\For{$q_2=1$ \textbf{to} $Q$}
		\State $\gamma^{(j,i)}:=\delta_2^{(q_1,g_1)}+\delta_2^{(q_2,g_2)}$
		\State $j:=j+1$
		\EndFor
		\EndFor
		\ElsIf{$n=4$}
		\For{$q_1=1$ \textbf{to} $Q$}
		\For{$q_2=1$ \textbf{to} $Q$}
		\State $\gamma^{(j,i)}:=\delta_2^{(q_1)}+\delta_2^{(q_2)}$
		\State $j:=j+1$
		\EndFor
		\EndFor
		\EndIf
		\EndFor
		\EndFor
		
		\EndFor
		\EndFor
		\For {$i = 1$ \textbf{to} $\binom{n}{n/2}$}
		\State $j:=1$
		\For {$m = 1$ \textbf{to} $MQ^2$ \textbf{with increments} $Q^2$} 
		\State $\Delta:=$MAX($\gamma^{(m,i)},\gamma^{(m+1,i)}$)
		\For {$q=2$ \textbf{to} $Q^2-1$}
		\State $\Delta:=$MAX($\Delta,\gamma^{(m+q,i)}$)
		\EndFor
		\State $\Gamma^{(j,i)}:=\Delta$
		\For {$z=1$ \textbf{to} $n/2$}
		\State $\hat{\mathcal{I}}:=\mathcal{I}^{(i,z)}$
		\State $\epsilon^{(z)}:=-|\mathbf{y}^{(\hat{\mathcal{I}})}|^{2}/N_0$
		\EndFor
		\State $\Gamma^{(j,i)}:=\Gamma^{(j,i)}+\sum_{z=1}^{n/2} \epsilon^{(z)}$
		\State $j:=j+1$
		\EndFor
		\EndFor
		\State $\Gamma:=\Gamma^{\mathrm{T}}$
	\end{algorithmic} 
\end{algorithm}

\begin{algorithm}
	\normalsize 
	\begin{algorithmic}[1]
		\setcounter{ALG@line}{52}
		\For {$v=1$ \textbf{to} $\binom{M}{2}$}
		\State $\Tilde{\mathcal{I}}:=$combinatorics\_n2c($v-1,2$) $+1$
		\For {$r=1$ \textbf{to} $\binom{n}{n/2}$}
		\State $\mathbf{\Lambda}^{(r,v)}:=\Gamma^{(r,\Tilde{\mathcal{I}}^{(1)})}+\Gamma^{(\binom{n}{n/2}-r+1,\Tilde{\mathcal{I}}^{(2)})}$
		\EndFor
		\EndFor
		\For {$a_1=0$ \textbf{to} $\binom{M}{2}-1$}
		\For {$a_2=0$ \textbf{to} $\binom{n}{n/2}-1$}
		\If {($a_1+\binom{M}{2}a_2\geq2^{p_1}$)}
		\State $\mathbf{\Lambda}^{(a_2+1,a_1+1)}:=-\infty$
		\EndIf
		\EndFor
		\EndFor
		\State $(\hat{a_2},\hat{a_1})=\underset{a_1,a_2}{\arg\: \max}\hspace{0.1cm}{\mathbf{\Lambda}^{(a_2,a_1)}}$ 
		\State $\hat{\mathbf{v}}:=$$\mathrm{combinatorics\_n2c}$($\hat{a_1}-1$,$2$)$+1$ 
		\State $\hat{\mathbf{u}}:=$$\mathrm{combinatorics\_n2c}$($\hat{a_2}-1$,$n/2$)$+1$
		\State $\hat{\mathbf{w}}:=$$\mathrm{combinatorics\_n2c}$($\binom{n}{n/2}-\hat{a_2}-1$,$n/2$)$+1$

		\If{$n=8$}
		
		\State {\scriptsize{$\hat{x}_1:=\underset{q}{\arg\:\min}{ \left \| [\mathbf{y}^{(\hat{\mathbf{u}}^{(1)})}\mathbf{y}^{(\hat{\mathbf{u}}^{(2)})}] - [\mathbf{c}^{(\hat{\mathbf{u}}^{(1)})}\mathbf{c}^{(\hat{\mathbf{u}}^{(2)})}] \mathcal{M}^{(\hat{\mathbf{v}}^{(1)},q)} \right \|}^2$}}
		\State {\scriptsize$\hat{x}_2:=\underset{q}{\arg\:\min}{ \left \| [\mathbf{y}^{(\hat{\mathbf{u}}^{(3)})}\mathbf{y}^{(\hat{\mathbf{u}}^{(4)})}] - [\mathbf{c}^{(\hat{\mathbf{u}}^{(3)})}\mathbf{c}^{(\hat{\mathbf{u}}^{(4)})}] \mathcal{M}^{(\hat{\mathbf{v}}^{(1)},q)} \right \|}^2$}
		\State {\scriptsize$\hat{x}_3:=\underset{q}{\arg\:\min}{ \left \| [\mathbf{y}^{(\hat{\mathbf{w}}^{(1)})}\mathbf{y}^{(\hat{\mathbf{w}}^{(2)})}] - [\mathbf{c}^{(\hat{\mathbf{w}}^{(1)})}\mathbf{c}^{(\hat{\mathbf{w}}^{(2)})}] \mathcal{M}^{(\hat{\mathbf{v}}^{(2)},q)} \right \|}^2$}
		\State {\scriptsize $\hat{x}_4:=\underset{q}{\arg\:\min}{ \left \| [\mathbf{y}^{(\hat{\mathbf{w}}^{(3)})}\mathbf{y}^{(\hat{\mathbf{w}}^{(4)})}] - [\mathbf{c}^{(\hat{\mathbf{w}}^{(3)})}\mathbf{c}^{(\hat{\mathbf{w}}^{(4)})}] \mathcal{M}^{(\hat{\mathbf{v}}^{(2)},q)} \right \|}^2$}
		\State $\hat{\mathbf{x}}=[\hat{x}_1\hspace{0.1cm}\hat{x}_2\hspace{0.1cm}\hat{x}_3\hspace{0.1cm}\hat{x}_4]$
		
		\ElsIf{$n=4$}
		
		\State {\scriptsize $\hat{x}_1:=\underset{q}{\arg\:\min}{ \left \| [\mathbf{y}^{(\hat{\mathbf{u}}^{(1)})}\mathbf{y}^{(\hat{\mathbf{u}}^{(2)})}] - [\mathbf{c}^{(\hat{\mathbf{u}}^{(1)})}\mathbf{c}^{(\hat{\mathbf{u}}^{(2)})}] \mathcal{M}^{(\hat{\mathbf{v}}^{(1)},q)} \right \|}^2$}
		\State {\scriptsize$\hat{x}_2:=\underset{q}{\arg\:\min}{ \left \| [\mathbf{y}^{(\hat{\mathbf{w}}^{(1)})}\mathbf{y}^{(\hat{\mathbf{w}}^{(2)})}] - [\mathbf{c}^{(\hat{\mathbf{w}}^{(1)})}\mathbf{c}^{(\hat{\mathbf{w}}^{(2)})}] \mathcal{M}^{(\hat{\mathbf{v}}^{(2)},q)} \right \|}^2$}
		\State $\hat{\mathbf{x}}=[\hat{x}_1\hspace{0.1cm}\hat{x}_2]$	
		\EndIf\newline
		\Return $\hat{a_1}, \hat{a_2}, \hat{\mathbf{x}}$
		
	\end{algorithmic} 
\end{algorithm}
At the receiver side, after removing the CP, de-interleaving and performing FFT,  ML detection can be considered to decode the $\alpha$th subblock. The subblock set, which includes all possible subblock realizations, is defined as
\begin{equation}
    \mathcal{S}=\{\mathbf{s}_1,\mathbf{s}_2,\dots\mathbf{s}_{2^p}\}.
    \label{Eq9}
\end{equation}
Then, ML detection rule for SuM-OFDM-IM is given as
\begin{equation}
    \hat{\mathbf{s}}=\underset{\mathbf{s}\in\mathcal{S}}{\arg\: \min}{ \left \|\mathbf{y}_F^{(\alpha)}-\mathbf{c}_F^{(\alpha)} \odot\mathbf{s} \right \|}^2,
    \label{Eq10}
\end{equation}
where $\mathbf{y}_F^{(\alpha)}$ and $\mathbf{c}_F^{(\alpha)}$ and  are the vectors of the received signals and the channel fading coefficients corresponding to $\alpha$th subblock, respectively, and $\odot$ is the element-wise product.  

The ML detection complexity per subblock is proportional to $2^{p}$, therefore, it is not practical and efficient for large values of $n$, $M$ and $Q$. Thus, we propose a novel LLR-based reduced  complexity ML detection algorithm for $n=4$ and $n=8$ as seen in Algorithm 1, which significantly reduces the detection complexity as discussed in Section V. Since we perform repetition coding for data symbols, we present a modified LLR detector. This detector achieves the same performance as the ML detector since it searches for all possible MAP and SAP combinations.

 The functions considered in Algorithm 1 are $\mathrm{MAX}(a,b)=\mathrm{max}(a,b)+\ln(1+e^{-|a-b|})$ and the combinatorial algorithm $\mathcal{I}=\mathrm{combinatorics\_n2c}(c,d)$ that takes the integer $c$ and the number of indices $d$ as inputs and outputs the set of indices $\mathcal{I}$ \cite{OFDMIM}. 

The major steps of Algorithm 1 are explained below:
\begin{enumerate}
	\item The inputs of the algorithm  are $\alpha$th received subblock $(\mathbf{y}_F^{(\alpha)})$, the channel fading coefficients corresponding to $\alpha$th subblock $(\mathbf{c}_F^{(\alpha)})$, noise power $(N_0)$, the set of all modes $(\mathcal{M})$, the number of modes $(M)$, modulation order $(Q)$, the number of bits conveyed by IM $(p_1)$ and the subblock size $(n)$.
	\item To determine the most likely MAP and SAP combination, LLR values are obtained between the lines (2-58). Specifically, between the lines (2-35), we provide a clever LLR calculation technique due to the unique symbol assignment strategy of our scheme. An iterative approach is applied between the lines (39-42) as in \cite{DUAL}.
	\item By assigning $-\xi$, $(\xi>>1)$ to LLR values corresponding to illegal MAP and SAP combinations between the lines (59-65), we avoid them. Thus, these illegal values can not be selected.
	\item Based on LLR values, MAP and SAP are jointly detected between the lines (66-69). For example, assuming $M=4, n=4$, we obtain
	\begin{equation}
		\mathbf{\Lambda}=
			\begin{bmatrix}
		\Lambda_{11} &\Lambda_{12}  &\Lambda_{13}  &\Lambda_{14}  &\Lambda_{15}  &\Lambda_{16} \\ 
		\Lambda_{21} &\Lambda_{22}  &\Lambda_{23}  &\Lambda_{24}  &\Lambda_{25}  &\Lambda_{26} \\ 
		\Lambda_{31} &\Lambda_{32}  &\Lambda_{33}  &\Lambda_{34}  &\Lambda_{35}  &\Lambda_{36} \\  
		\Lambda_{41} &\Lambda_{42}  &\Lambda_{43}  &\Lambda_{44}  &\Lambda_{45}  &\Lambda_{46} \\ 
		\Lambda_{51} &\Lambda_{52}  &\Lambda_{53}  &\Lambda_{54}  &\Lambda_{55}  &\Lambda_{56} \\ 
		\Lambda_{61} &\Lambda_{62}  &-\xi  &-\xi  &-\xi &-\xi \\ 
		\end{bmatrix},
		\label{Eq11}
	\end{equation}
	where $\Lambda_{ij}$ is the LLR value in $i$th row and $j$th column of the LLR matrix $\mathbf{\Lambda}$, which is formed between the lines (53-58). Columns and rows of $\mathbf{\Lambda}$ represents the possible MAPs and SAPs, respectively. As seen from (\ref{Eq11}), $\mathbf{\Lambda}$ includes LLR values for all possible MAP and SAP combinations; therefore, this detector provides the same error performance as the ML detector. 
	\item After detecting the active MAP and SAP, data symbols are decoded between the lines (70-80).
	\item Finally, algorithm outputs the decoded data symbols $(\hat{\mathbf{x}})$ and decimal equivalent of IM bits  $(\hat{d}=\hat{a}_1+\binom{M}{2}\hat{a}_2)$.
	
\end{enumerate}

 { \textit{Remark (Generalization):} To generalize the LLR-based detector for arbitrary $n$, only the following changes should be performed in Algorithm 1:
	\begin{itemize}
		\item The condition in line 17 is changed as $n \neq 4$ rather than $n = 8$.
		\item The lines between (70-80) are modified. For any values of $n$, a total of $n/2$ metric calculations are performed. If $\kappa\le n/4$, the parameters $\hat{u}^{(2\kappa-1)}$, $\hat{u}^{(2\kappa)}$, and $\hat{v}^{(1)}$ are exploited to detect $\hat{x}_\kappa$, $\kappa=1,\cdots,n/2$. If $\kappa>n/4$, the parameters $\hat{w}^{(2(\kappa-n/4)-1)}$, $\hat{w}^{(2(\kappa-n/4))}$, and $\hat{v}^{(2)}$ are exploited to detect $\hat{x}_\kappa$.
	\end{itemize}
}

\section{Mode Selection for SuM-OFDM-IM} 

In this section, we discuss our strategy for mode selection.  According to \cite{MMOFDMIM}, the optimal modes must maximize the minimum intra-mode distance (MIAD)
\begin{align}
    \epsilon_1&=\underset{\iota,\kappa \in\{1,\dots,2^p\}}{\min}
    { \left \| \mathbf{S}_{\iota} - \mathbf{S}_{\kappa} \right \|}^2_F, \nonumber  \\
    \text{s.t.} & \hspace{0.2cm} E\{\|\mathbf{S} \|^2_F\}=n \:\:\text{and}\:\: \mathrm{rank}(\mathbf{S}_{\iota}- \mathbf{S}_{\kappa})=1,
    \label{Eq12}
\end{align}
where $\mathbf{S}=\mathrm{diag}(\mathbf{s})$, $\mathrm{diag}(\mathbf{a})$ represents a diagonal matrix whose diagonal elements are $\mathbf{a}$, and the minimum inter-mode distance (MIRD)
\begin{figure}[!t]
	\centering
	\vspace{-0.4cm}
	\includegraphics[width=0.75\linewidth, height=5.0cm ]{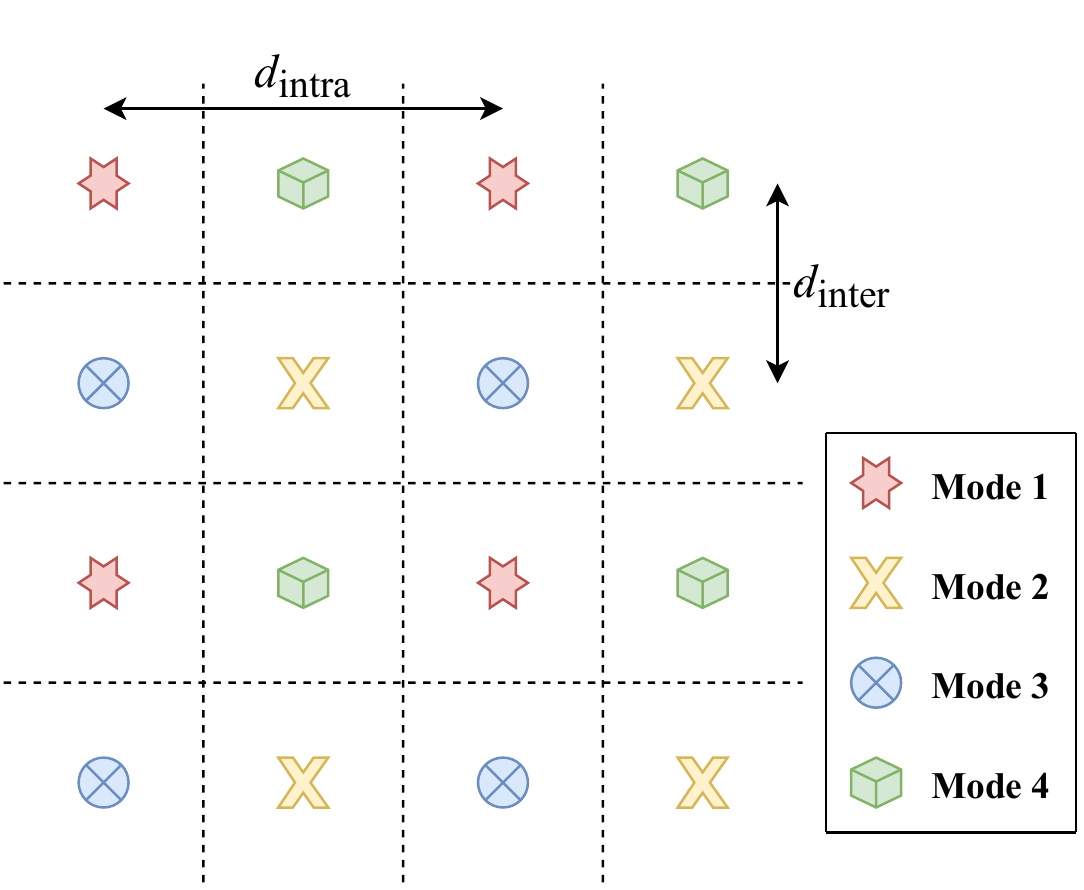}	
	\vspace{-0.2cm}
	\caption{Partition of $16$-QAM constellation for $M=4$ and $Q=4$.}
	\vspace{-0.6cm}
	\label{Fig. 1}
\end{figure}
\begin{align}
    \epsilon_2&=\underset{\iota,\kappa \in\{1,\dots,2^p\}}{\min}
    { \left \| \mathbf{S}_{\iota} - \mathbf{S}_{\kappa} \right \|}^2_F, \nonumber  \\
    \text{s.t.} & \hspace{0.2cm} E\{\|\mathbf{S} \|^2_F\}=n \:\:\text{and}\:\: \mathrm{rank}(\mathbf{S}_{\iota}- \mathbf{S}_{\kappa})=2,
    \label{Eq13}
\end{align}
where $\mathbf{S}_{\iota}$ and $\mathbf{S}_{\kappa}$ are two different realizations of $\mathbf{S}$. For simplicity, we have removed superscript $(\alpha)$ in $\mathbf{S}^{(\alpha)}$, since all subblocks are identical. The modes can be created by partitioning PSK/QAM constellations as in \cite{MMOFDMIM} and it is proven that QAM constraint performs better than PSK constraint for $MQ>4$. Therefore, we focus on only QAM constellation for our system model: 
\begin{itemize}
    \item QAM constraint to maximize MIRD:
\end{itemize}
\begin{equation}
    d_{\mathrm{inter}}(M,Q)= \begin{cases}
    2\sqrt{\frac{6}{5QM-4}}, & \text{for rectangular $MQ-$QAM}\\
    \sqrt{\frac{6}{QM-1}}, & \text{for square $MQ-$QAM}
    \end{cases} 
    \label{Eq14}
\end{equation}
\begin{itemize}
    \item QAM constraint to maximize MIAD:
\end{itemize}
\begin{align}
 &d_{\mathrm{intra}}(M,Q) = \nonumber  \\ &\begin{cases} 
     d_{\mathrm{inter}}(M,Q) \frac{\sqrt{5M}}{2}, & \text{for rectangular $MQ-$QAM ($Q=2$)}\\
     d_{\mathrm{inter}}(M,Q)\sqrt{M}, & \text{otherwise}
    \end{cases} 
    \label{Eq15}
\end{align}
\begin{itemize}
    \item QAM constraint to maximize MIAD for $MQ>>1$:
\end{itemize}
\begin{align}
     &d_{\mathrm{intra}}(M,Q) \approx d_{\mathrm{intra}}(Q)=   \nonumber  \\  &\begin{cases} 
    2\sqrt{\frac{6}{5Q}}, & \text{for rectangular $MQ-$QAM ($Q\neq2$)}\\
    \sqrt{\frac{6}{Q}}, & \text{otherwise}
    \end{cases} 
    \label{Eq16}
\end{align}
with these constraints, modes can be obtained by partitioning QAM constellations for given parameters $(M,Q)$. For example, for $M=4,Q=4$, the optimal modes for $16$-QAM can be selected as in Fig. 3, where $d_{\mathrm{inter}}=0.6325$  and $d_{\mathrm{intra}}=1.2649$ from (\ref{Eq14}) and (\ref{Eq15}), respectively.

\section{Performance Analysis}
In this section, we derive an upper bound on the BER of the SuM-OFDM-IM system employing ML detection. From (\ref{Eq10}), the conditional pairwise error probability (CPEP), which is defined as the probability of  transmitting $\mathbf{S}$ and erroneously detecting $\hat{\mathbf{S}}$ conditioned on $\mathbf{c}_F$, is given as:
\begin{equation}
    \text{Pr}(\mathbf{S} \rightarrow \hat{\mathbf{S}}|\mathbf{c}_F)
    =\mathcal{Q}\bigg(\sqrt{\frac{\rho}{2} \Big\|(\mathbf{S}-\hat{\mathbf{S}})\mathbf{c}_F \Big\|^2 }\bigg).
    \label{Eq17}
\end{equation}
Then, by approximating $\mathcal{Q}(x)\approx \frac{e^{-x^2/2}}{12}+\frac{e^{-2x^2/3}}{4}$, the unconditioned PEP (UPEP) can be obtained as in \cite{OFDMIM}:
\begin{align}
    \text{Pr}\big(\mathbf{S} \rightarrow \hat{\mathbf{S}}\big) 
    &=E_{\mathbf{c}_F}\{\text{Pr}(\mathbf{S} \rightarrow \hat{\mathbf{S}}|\mathbf{c}_F)\} \nonumber\\
    &\approx\frac{1/12}{\text{det}(\mathbf{I}_n+\rho_1\mathbf{A})}+
    \frac{1/4}{\text{det}(\mathbf{I}_n+\rho_2\mathbf{A})},
    \label{Eq18}
\end{align}
where $\rho_1=1/(4N_{0}), \rho_2=1/(3N_{0})$, $\mathbf{A}=(\mathbf{S}-\hat{\mathbf{S}})^{\mathrm{H}}(\mathbf{S}-\hat{\mathbf{S}})$ and $(.)^{\mathrm{H}}$ denotes Hermitian transposition. Finally, according to the union bounding technique, the BER of SuM-OFDM-IM can be upper bounded by 
\begin{equation}
    P_e \leq \frac{1}{p2^p}\sum_{\mathbf{S}} \sum_{\hat{\mathbf{S}}} \text{Pr} \big(\mathbf{S} \rightarrow \hat{\mathbf{S}}\big) e\big(\mathbf{S},\hat{\mathbf{S}}\big),
    \label{Eq19}
\end{equation}
where Pr$(\mathbf{S} \rightarrow \hat{\mathbf{S}})$ is given by (\ref{Eq18}) and $e(\mathbf{S},\hat{\mathbf{S}})$ represents the number of bit errors for the corresponding pairwise error event.

{ We have $\text{det}(\mathbf{A})=\prod_{\zeta=1}^{r} \lambda_{\zeta}(\mathbf{A})$ where $\lambda_{\zeta}(\mathbf{A})$ is the $\zeta$th eigenvalue of $\mathbf{A}$ and $r=\mathrm{rank}(\mathbf{A})$ determines the diversity order of the system, since $P_e$ decays with $r$. At high SNR values, by ignoring $\mathbf{I}_n$ terms in (\ref{Eq18}), (\ref{Eq19}) can be simplified as follows:}
\begin{equation}
P_e \leq \frac{\rho_1^{-r} + 3\rho_2^{-r}}{12p2^p}\sum_{\mathbf{S}} \sum_{\hat{\mathbf{S}}} \Big(\prod_{\zeta=1}^{r} \lambda_{\zeta}(\mathbf{A})\Big)^{-1}  e\big(\mathbf{S},\hat{\mathbf{S}}\big).
\label{Eq20}
\end{equation}
 We investigate two cases: $(i)$ erroneous detection of index bits and $(ii)$ erroneous detection of a single or multiple data symbols while the index bits are decoded correctly. For case $(i)$, due to our novel MAP and SAP selection strategy, we always obtain $r\ge2$. For example, from Table I, assume that $\mathbf{v}=[1,2]$ is selected and $\hat{\mathbf{v}}=[1,3]$ is decoded erroneously, which is the worst case error event due to overlapping index values of $\mathbf{v}$ and $\hat{\mathbf{v}}$. Even in this error event, $r$ is obtained as $2$ as in plain OFDM-IM systems \cite{OFDMIM}. If $\mathbf{v}$ and $\hat{\mathbf{v}}$ have non-overlapping indices, $r$ might be even equal to $3$ or $4$. To explain case $(ii)$, we illustrate the following example. Assume that $n=4$, $M=4$, $Q=4$ and $\mathbf{S}=\mathrm{diag}([\chi_{11}\:\chi_{11}\:\chi_{31}\:\chi_{31}])$, which is obtained as in (3), is transmitted and $\hat{\mathbf{S}}=\mathrm{diag}([\chi_{11}\:\chi_{11}\:\chi_{32}\:\chi_{32}])$ is erroneously detected, i.e, MAP and SAP are detected correctly; however, the second symbol is decoded erroneously $(\chi_{31}\rightarrow\chi_{32})$. Then, we obtain:
\begin{equation}
\mathbf{A}=
	\begin{bmatrix}
	0 &0  &0    &0 \\ 
	0 &0  &0    &0 \\ 
	0 &0  &1.6  &0 \\ 
	0 &0  &0    &1.6 
	\end{bmatrix},
	\label{Eq21}
\end{equation}
 where $r=\mathrm{rank}(\mathbf{A})=2$. As seen from this example, due to clever symbol assignment strategy of our scheme, even the erroneous detection of a single data symbol ensures a diversity order of $2$. It can be easily shown that for all error events of this type, we again obtain at least $r=2$. 
 
 \begin{table}[t]
 	\centering
 	\caption{Percentage of Error Events}	
 	\renewcommand{\arraystretch}{1.4}
 	\begin{tabular}{|>{\centering\arraybackslash}m{1.2cm}  |>{\centering\arraybackslash}m{1.8cm}|>{\centering\arraybackslash}m{1.8cm}|>{\centering\arraybackslash}m{1.8cm}|}	
 		\hline
 		\textbf{$(M,Q)$} & $r=2$ & $r=3$ & $r=4$		\\ \hhline{====}
 		$(4,4)$  & $4.79\%$  & $15.07\%$ & $80.14\%$    \\ \hline
 		$(8,2)$  & $5.10\%$  & $14.95\%$ & $79.95\%$    \\ \hline
 		$(4,16)$ & $1.14\%$  & $4.02\%$  & $94.84\%$	\\ \hline
 		$(16,4)$ & $1.27\%$  & $3.92\%$  & $94.81\%$	\\ \hline
 	\end{tabular}
 \end{table}

To provide further insights on the performance of SuM-OFDM-IM, we obtain the percentage of error events with $r=2,3$ and $4$  in Table II for varying $(M,Q)$ by considering all possible error events for $n=4$. It should be noted that the percentage of error events has a remarkable affect on the overall error performance of a system from (\ref{Eq19}). As it can be deduced from Table II, since $r=4$ is obtained much more frequently than other possible rank values, our proposed system is capable of providing a superior error performance. It can be seen from Table II that as the overall constellation size $(MQ)$ increases, the percentage of $r=4$ increases due to the increase in the number of data symbols. Because, when MAP and SAP are detected correctly and two data symbols are decoded erroneously, we obtain $r=4$. Finally, in light of the above discussion, we obtain 
\begin{equation}
\underset{\mathbf{S},\hat{\mathbf{S}}}{\min} \ \mathrm{rank}(\mathbf{A})=2,
\label{Eq22}
\end{equation}
 which proves that our system provides a diversity order of $2$.

\section{Simulation Results} 

\begin{figure}[!t]
	\centering
	\includegraphics[width=0.8\linewidth ]{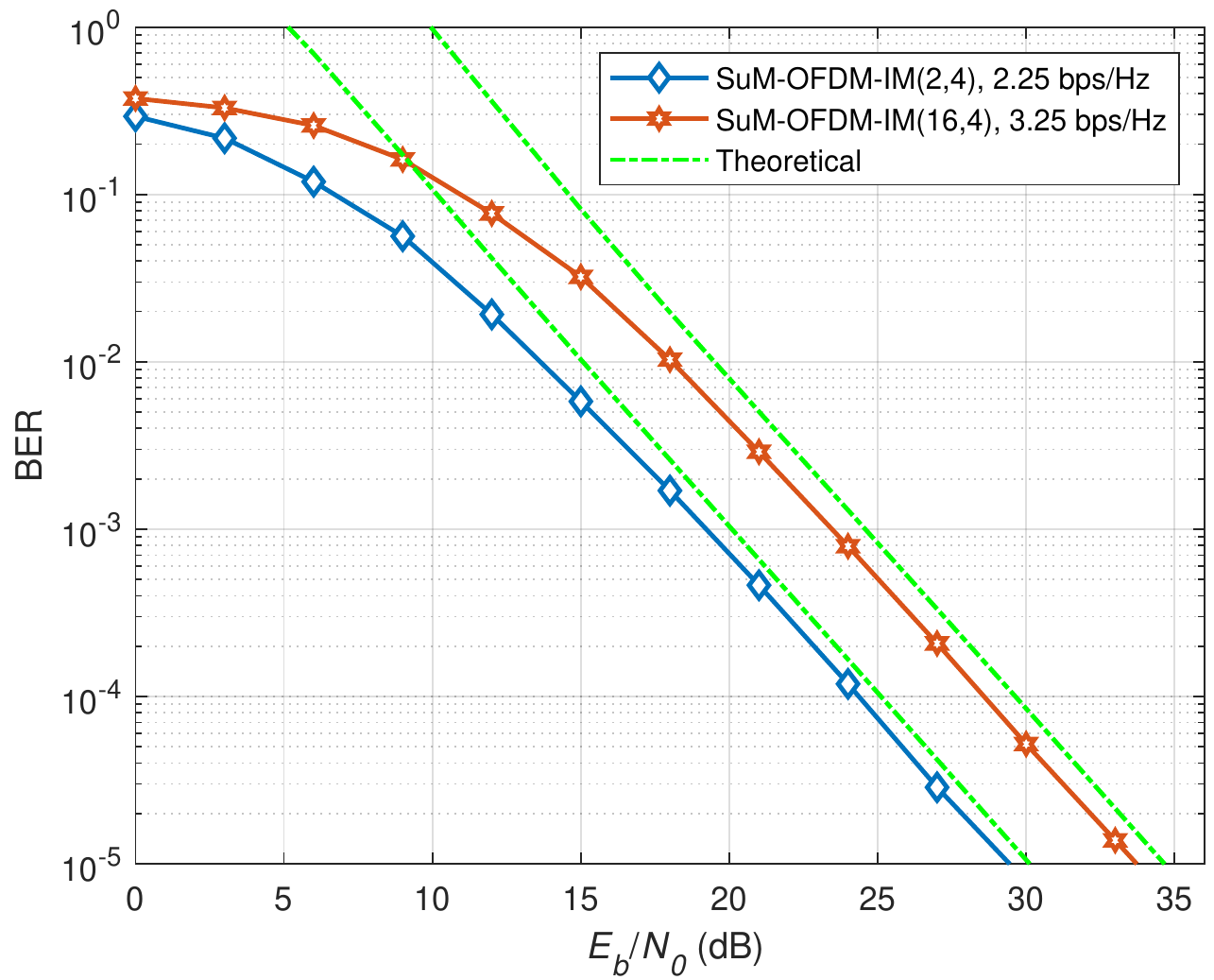}	
	\caption{Theoretical BER upper bound of SuM-OFDM-IM with simulation results.}
	\vspace{-0.2cm}
	\label{Fig. 3}
	\vspace{-0.3cm}
\end{figure}

In this section, we perform Monte Carlo simulations to compare the error performance of SuM-OFDM-IM with reference schemes. All simulations have been performed under frequency-selective Rayleigh fading channels and perfect channel estimation is assumed at the receiver side. OFDM parameters are adjusted to $N=128$, $v=10$ and $L=16$ for all simulations. We define "OFDM-IM$(n,k)$ and CI-OFDM-IM$(n,k)$"   as $k$ out of $n$ subcarriers being active in a subblock, "DM-OFDM$(n,k)$" as $k$ subcarriers out of $n$ exploiting the primary $M$-ary PSK/QAM constellation, "MM-OFDM-IM$(Q,M)$, SuM-OFDM-IM$(Q,M)$, and {S-SuM-OFDM-IM$(Q,M)$}" with $M$ modes, each including $Q$ symbols.

As shown in Fig. 4, the theoretical BER curve obtained by (\ref{Eq19}) is an accurate upper bound for simulation results of SuM-OFDM-IM for varying parameters $(Q,M)$. {Since (\ref{Eq18}) is an approximation, as spectral efficiency  increases, the number of terms in (\ref{Eq19}) also increases; therefore, the gap between simulation and theoretical curves does not diminish.} 

\begin{figure}[!t]
	\centering
	\includegraphics[width=0.8\linewidth ]{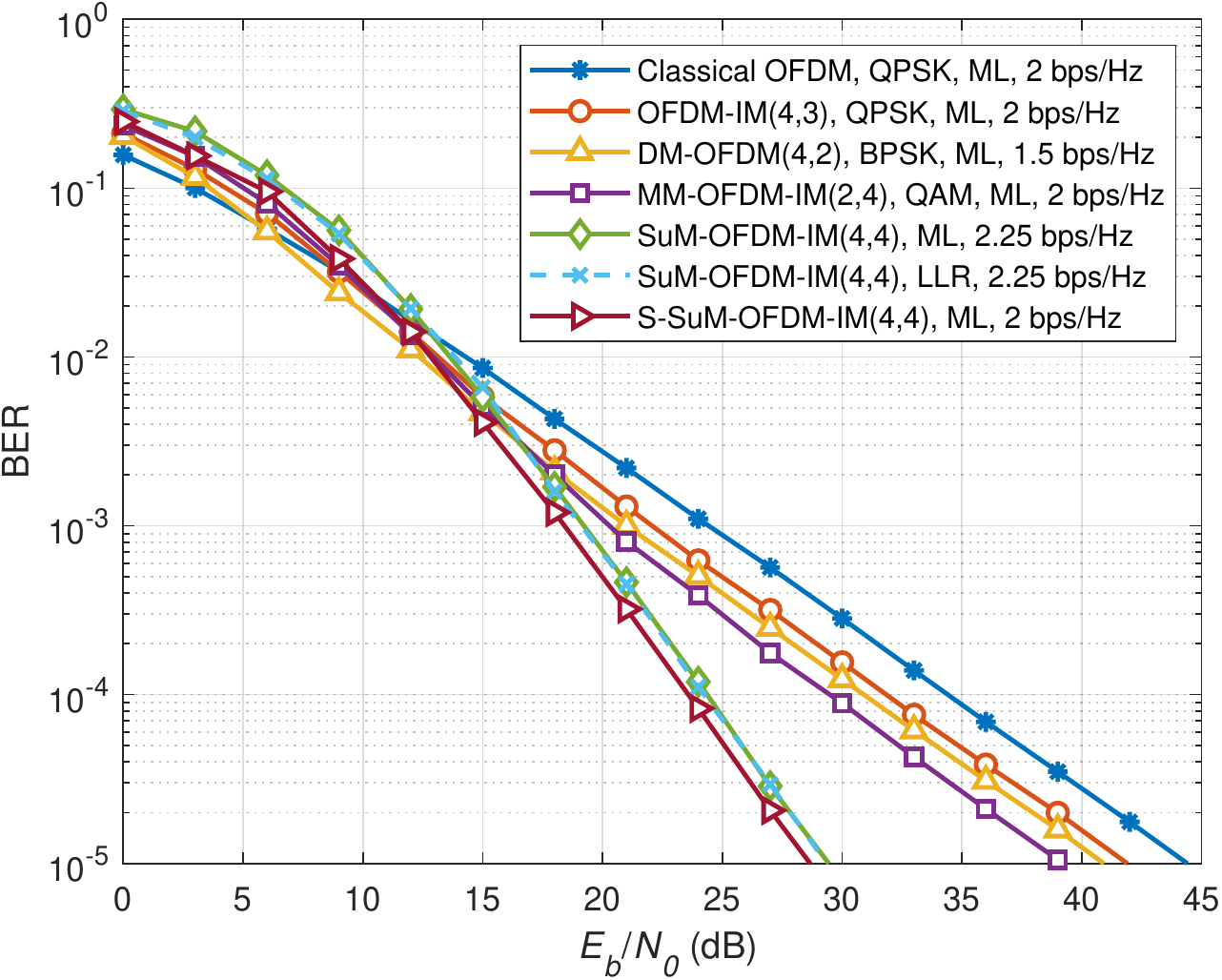}	
	\vspace{-0.2cm}
	\caption{{Error performance comparison of SuM-OFDM-IM and S-SuM-OFDM-IM with reference systems.}}
	\vspace{-0.4cm}
	\label{Fig. 4}
\end{figure}
{It is observed from Fig. 5 that SuM-OFDM-IM and S-SuM-OFDM-IM have superior BER performances over reference systems for approximately the same spectral efficiency. As seen from Fig. 5, SuM-OFDM-IM and S-SuM-OFDM-IM outperform classical OFDM, OFDM-IM, DM-OFDM and MM-OFDM-IM.} Additionally, at a BER value of $10^{-5}$, SuM-OFDM-IM provides almost $10$ dB gain over MM-OFDM-IM although providing a $12.5\%$ higher spectral efficiency. It is also shown that LLR-based reduced ML detector provides the same error performance as that of the ML detector. {Although SuM-OFDM-IM performs only slightly worse in terms of error performance than S-SuM-OFDM-IM, its spectral efficiency is $12.5\%$ higher.}

\begin{figure}[!t]
	\centering
	\includegraphics[width=0.8\linewidth ]{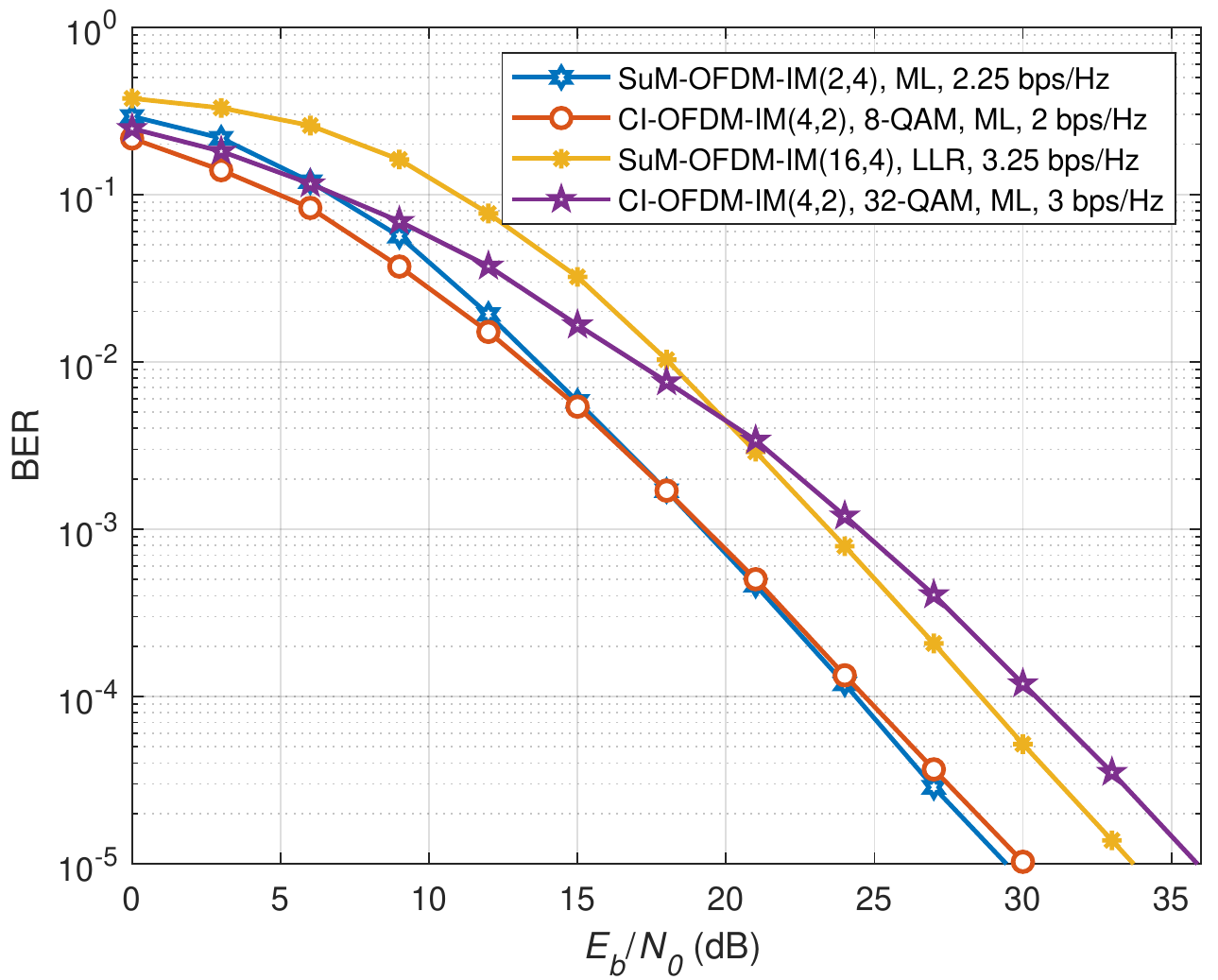}	
	\vspace{-0.2cm}
	\caption{Error performance comparison of SuM-OFDM-IM with CI-OFDM-IM.}
	\vspace{-0.4cm}
	\label{Fig. 5}
\end{figure}
In Fig. 6, SuM-OFDM-IM is compared with CI-OFDM-IM, which also provides a second order diversity gain. First, we consider $2$ and $2.25$ bps/Hz for CI-OFDM-IM and SuM-OFDM-IM, respectively. SuM-OFDM-IM provides a marginal gain at a BER value of $10^{-5}$ than CI-OFDM-IM  despite a $12.5\%$ improvement in terms of spectal efficiency. Moreover, we consider $3$ and $3.25$ bps/Hz for CI-OFDM-IM and SuM-OFDM-IM, respectively. SuM-OFDM-IM is able to provide a noticeable gain at a BER value of $10^{-5}$ than CI-OFDM-IM  despite yielding a $8.33\%$ higher spectral efficiency. It can be clearly concluded that the performance difference between SuM-OFDM-IM and CI-OFDM-IM increases with spectral efficiency.

\begin{figure}[t]
	\centering
	\includegraphics[width=0.8\linewidth ]{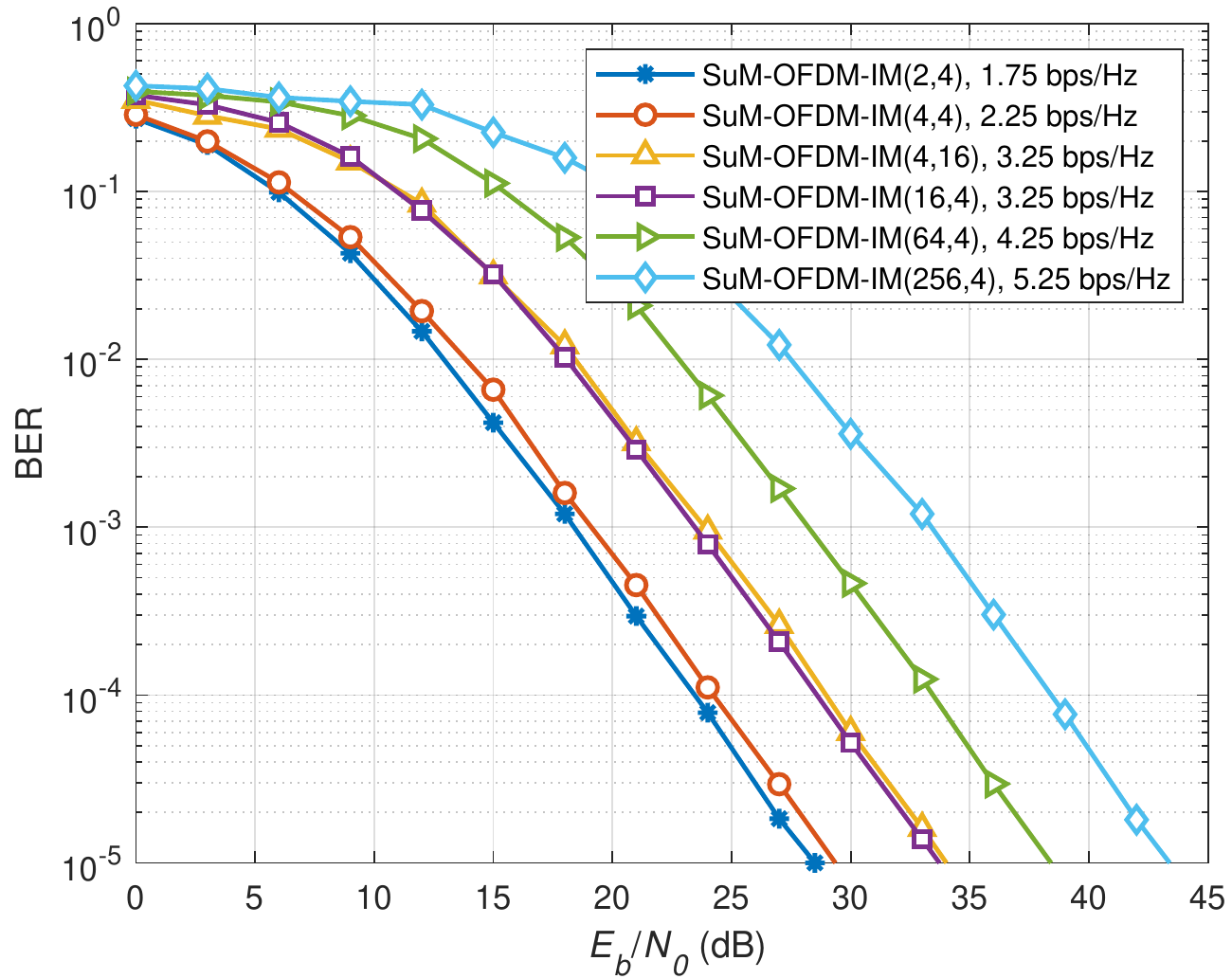}	
	\vspace{-0.2cm}
	\caption{Error performance of SuM-OFDM-IM for different spectral efficiencies with LLR detector for $n=4$.}
	\vspace{-0.4cm}
	\label{Fig. 6}
\end{figure}
In Fig. 7, the error performance of SuM-OFDM-IM with varying $(Q,M)$ parameters is given for different spectral efficiency values and $n=4$. LLR-based reduced complexity ML detector is employed in all cases. As seen from Fig. 7, the slopes of the curves are equal and the error performance degrades with increasing spectral efficiency. 

\begin{figure}[t]
	\centering
	\includegraphics[width=0.8\linewidth ]{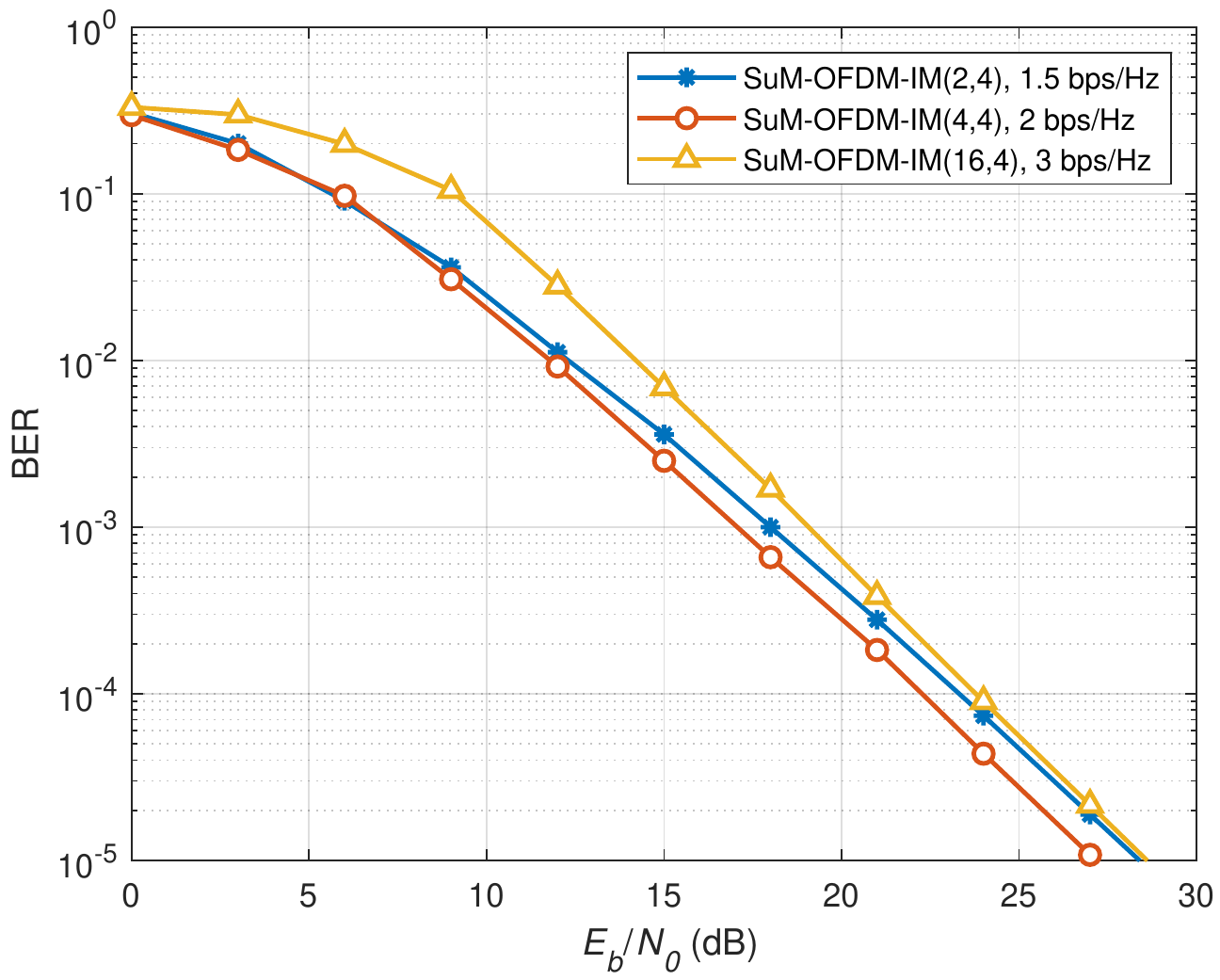}	
	\vspace{-0.2cm}
	\caption{Error performance of SuM-OFDM-IM for different spectral efficiencies with LLR detector for $n=8$.}
	\vspace{-0.4cm}
	\label{Fig. 7}
\end{figure}
Error performance of SuM-OFDM-IM for varying parameters $(Q,M)$ is shown in Fig. 8 where $n=8$. We observe that the error performance is inversely proportional to  the size of the constellation. As $Q$ increases, the percentage of higher rank values increases and it can be seen from Fig. 8 that at a BER value of $10^{-5}$, the performances of schemes with $2$ bps/Hz and $3$ bps/Hz are close to each other. By comparing Figs. 7 and 8 to investigate the effect of $n$ on the BER performance, it can readily be deduced that the system with $n=8$ and $3$ bps/Hz outperforms the system with $n=4$ and $3.25$ bps/Hz. This is because $r$ may even be equal to $8$ when $n=8$, resulting in an improvement in error performance; nevertheless, the complexity increases proportional to $n$ as seen in Table III. Additionally, it is more challenging to reach a higher spectral efficiency for $n=8$ compared to $n=4$ by (\ref{Eq7}). Therefore, we introduce interesting trade-offs among spectral efficiency, decoding complexity and error performance.

\begin{figure}[!t]
	\centering
	\includegraphics[width=0.8\linewidth ]{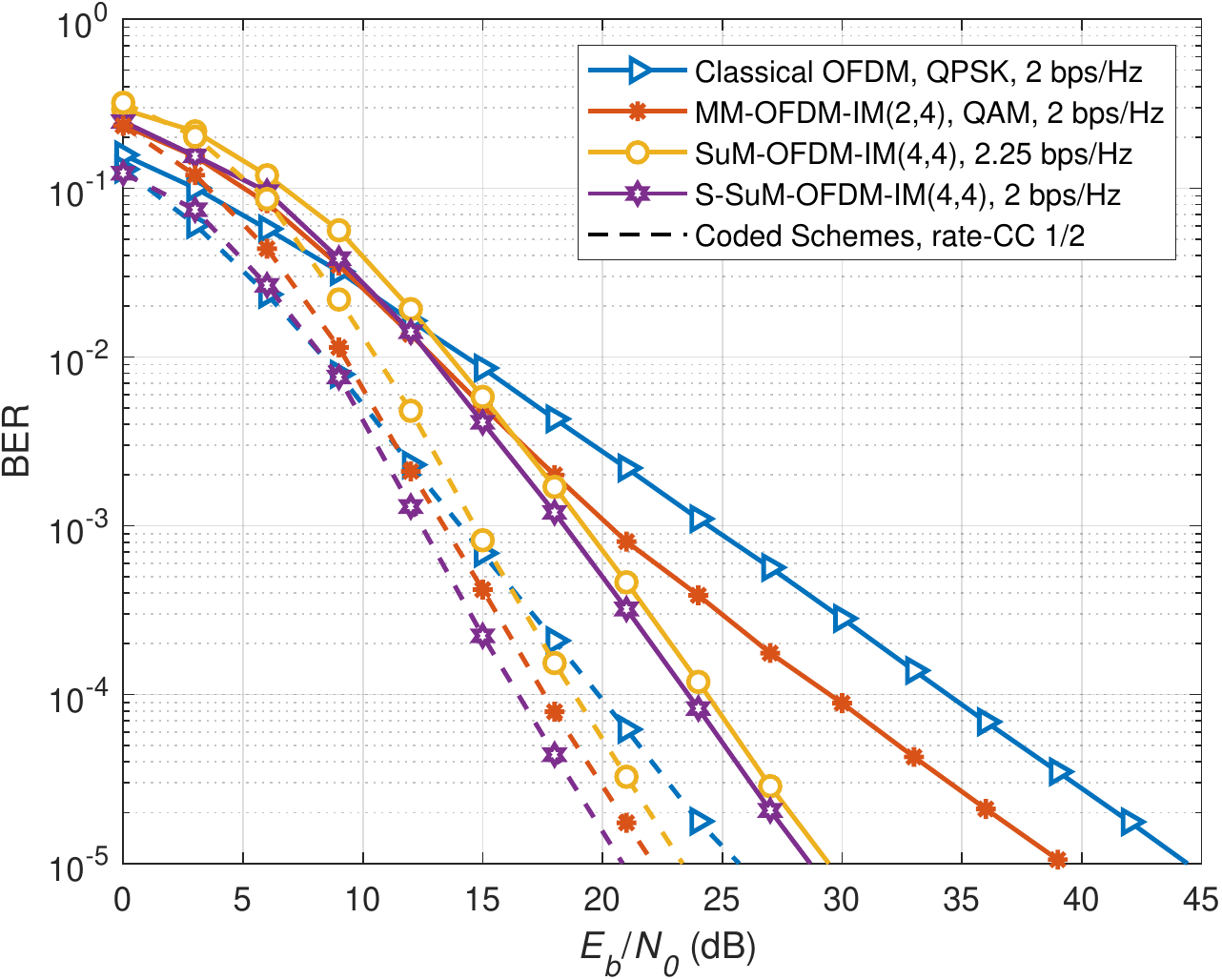}	
	\vspace{-0.2cm}
	\caption{{Uncoded/Coded performance of OFDM, MM-OFDM-IM, SuM-OFDM-IM, and S-SuM-OFDM-IM.}}
	\label{Fig. 8}
	\vspace{-0.3cm}
\end{figure}

{In Fig. 9, the error performance curves of uncoded and coded OFDM, MM-OFDM-IM, SuM-OFDM-IM, and S-SuM-OFDM-IM are given.} For all coded schemes, the rate-$1/2$ convolutional code (CC) with octal generator sequence of [1, 3] and a bit-interleaver are employed. {As seen from Fig. 9, SuM-OFDM-IM and S-SuM-OFDM-IM outperform OFDM at a BER value of $10^{-5}$ also with channel coding.} Moreover, it is observed that the difference between coded SuM-OFDM-IM and coded OFDM increases for higher SNR values due to different slopes of their curves. While the advantage of SuM-OFDM-IM is not significant compared to MM-OFDM-IM for this specific case, SuM-OFDM-IM can also be a candidate for potential future uncoded applications due to its diversity gain. {However, in the presence of coding, S-SuM-OFDM-IM outperforms MM-OFDM-IM. Therefore, we present an interesting trade-off between error performance and spectral efficiency for SuM-OFDM-IM and S-SuM-OFDM-IM.}

\begin{table}[t]
	\centering
	\caption{Decoding Complexity Comparison}	
	\renewcommand{\arraystretch}{1.4}
	\begin{tabular}{|>{\centering\arraybackslash}m{2.5cm} |>{\centering\arraybackslash}m{2cm}|>{\centering\arraybackslash}m{2.5cm}|}	
		\hline
		\textbf{System} & \textbf{Detector} & \textbf{Complexity Order} \\ \hline 
		OFDM             & ML                & $\mathcal{O}(Q)$       \\ \hline
		OFDM-IM  \cite{OFDMIM}      & Red. Comp. ML        & $\mathcal{O}(Q)$       \\ \hline
		DM-OFDM  \cite{DUAL}        & Suboptimal        & $\mathcal{O}(Q_A+Q_B)$ \\ \hline
		MM-OFDM-IM \cite{MMOFDMIM}  & Suboptimal        & $\mathcal{O}(\frac{Qn}{2}+\frac{Q}{2})$  \\ \hline
		CI-OFDM-IM \cite{CIOFDMIM}  & Red. Comp. ML        & $\mathcal{O}( Q^2 )$ \\ \hline
		SuM-OFDM-IM      & Red. Comp. ML        & $\mathcal{O}(\binom{n}{n/2}\frac{QM}{2})$ \\ \hline
		{S-SuM-OFDM-IM}  & {Red. Comp. ML}        & ${\mathcal{O}(\binom{n}{n/2}\frac{QM}{2})}$ \\ \hline
	\end{tabular}
\vspace{-0.4cm}
\end{table}

\begin{table}[t]
	\centering
	\caption{A Numerical Example For Complexity Comparison}	
	\renewcommand{\arraystretch}{1.3}
	\begin{tabular}{|>{\centering\arraybackslash}m{2cm} |>{\centering\arraybackslash}m{2cm}|>{\centering\arraybackslash}m{2.6cm}|}	
		\hline
		\textbf{System} & \textbf{Detector} & \textbf{CMs per subcarrier} \\ \hline 
		OFDM         & ML  & $4$       \\ \hline
		OFDM-IM      & Red. Comp. ML  & $4$       \\ \hline
		DM-OFDM      & Suboptimal  & $4$ \\ \hline
		MM-OFDM-IM   & Suboptimal  & $5$  \\ \hline
		CI-OFDM-IM   & Red. Comp. ML  & $64$ \\ \hline
		SuM-OFDM-IM  & Red. Comp. ML  & $48$ \\ \hline
		SuM-OFDM-IM  & ML  & $128$ \\ \hline
	\end{tabular}
\end{table}

\subsection{Complexity Analysis}

In this subsection, we analyze the decoding complexity order of SuM-OFDM-IM and make comparisons with the reference systems. A comparison for complexity per subcarrier in terms of complex multiplications (CMs), is given in Table III, where $Q_A$ and $Q_B$ stand for the primary and secondary constellations for DM-OFDM, respectively, and $k$ is the number of active subcarriers in a CI-OFDM-IM subblock. {Due to space limitation, we do not give the design of LLR-based reduced complexity ML detection for S-SuM-OFDM-IM, however, it can be designed as in Algorithm 1 in a similar manner. In this case, the complexity order of SuM-OFDM-IM and S-SuM-OFDM-IM would be equivalent as seen Table III.}

To further elaborate on the complexity order of the proposed system and reference systems, we give a numerical example in Table IV. For a fair comparison, the parameters in Figs. 6 and 5 are considered for CI-OFDM-IM with $2$ bps/Hz and other systems, respectively. As seen from Table IV, we provide an interesting trade-off among complexity and error performance. With these specific parameters, although SuM-OFDM-IM is the second most complex system, it provides a superior error performance as seen in Figs. 5 and 6. It should also be noted that the LLR-based reduced complexity ML detector dramatically reduces the complexity of SuM-OFDM-IM when compared to the ML detector. {Moreover, the LLR-based and low-complexity ML detector is not significantly complex for varying $n$, $Q$ and $M$ values. For example, for a spectral efficiency of $1.75$ bps/Hz, only $24$ CMs per subcarrier are required. Hence, SuM-OFDM-IM is applicable in practice with the proposed low-complexity detector.}
\vspace{-0.2cm}
\begin{figure}[!t]
\centering
	\includegraphics[width=0.82\linewidth ]{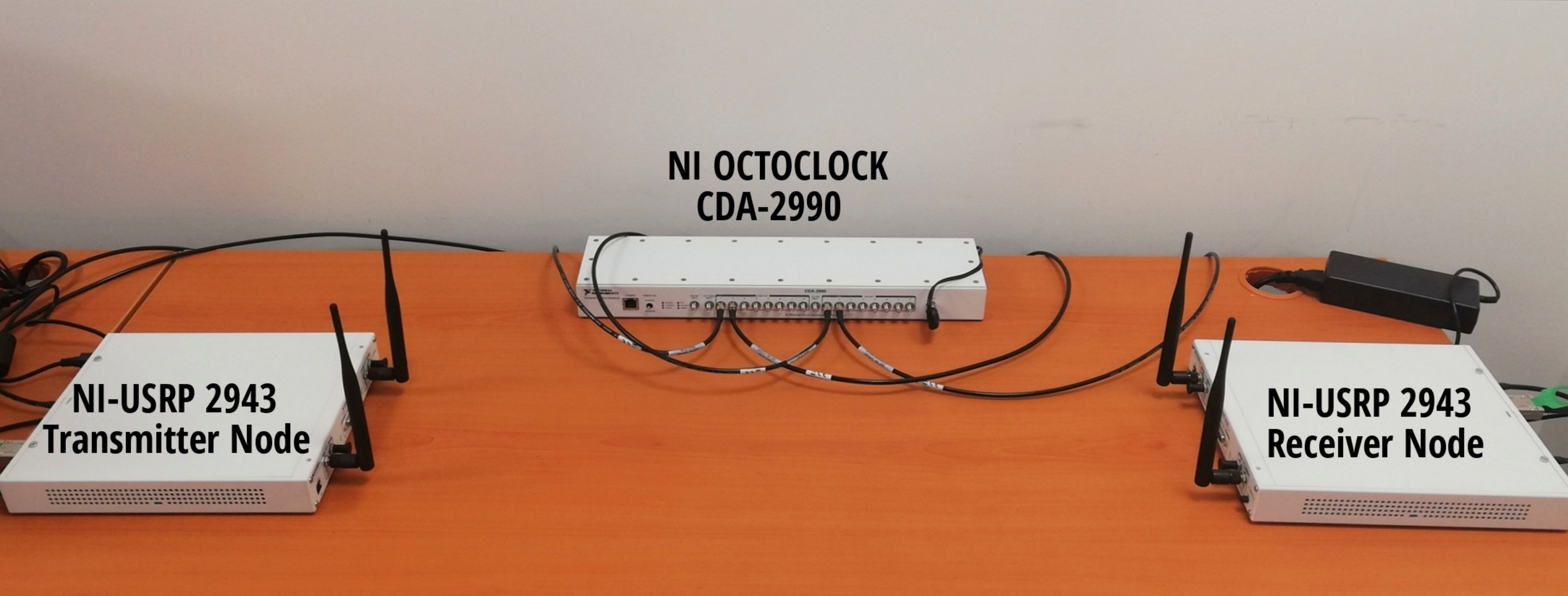}	
	\vspace{-0.2cm}
	\caption{Experimental testbed with USRP modules.}
	\label{Fig. 8}
	\vspace{-0.4cm}
\end{figure}

\section{Experimental Results} 
In this section, we evaluate the performance of SuM-OFDM-IM in a practical setup and give the models of the transmitter and receiver testbeds. As seen from Fig. 10, we use two USRP 2943 modules for transmitter and receiver nodes and an OctoClock CDA-2990 that generates $10$ MHz clock signal for synchronization. For programming, virtual instrument (VI) components of LabVIEW are utilized.

\vspace{-0.4cm}

\subsection{Transmitter Details}
\subsubsection{Channel Estimation}
Comb-type channel estimation method with one dimensional linear interpolation is employed to estimate the channel as in \cite{PRACT}. A certain number of subcarriers are assigned as pilot tones. In this experiment, we consider the ratio of data to pilot subcarriers as $4$, i.e, for $4$ data subcarriers, one pilot tone is employed.

\begin{figure}[!t]
	\centering
	\includegraphics[width=0.75\linewidth ]{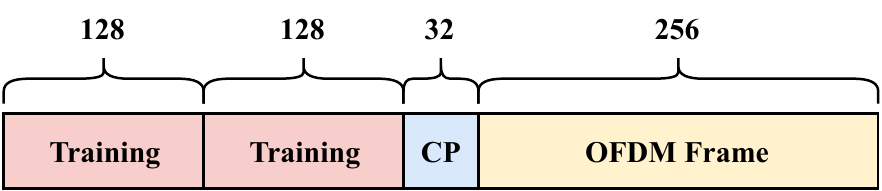}	
	\vspace{-0.2cm}
	\caption{Overall symbol structure.}
	\label{Fig. 9}
	\vspace{-0.4cm}
\end{figure}

\subsubsection{Symbol Structure}
 \begin{table}[t]
	\centering
	\caption{Experimental Parameters}	
	\renewcommand{\arraystretch}{1.4}
	\begin{tabular}{|>{\centering\arraybackslash}m{3.4cm}  |>{\centering\arraybackslash}m{3.4cm}|}	
		\hline
		
		Carrier Frequency  & $2.45$ GHz    \\ \hline
		I/Q data rate  & $5 \times 10^5$ samples/sec    \\ \hline
		Sampling rate $(f_s)$ &$6.25 \times 10^5$ samples/sec     \\ \hline
		Number of data subcarriers & $125$    \\ \hline
		Number of pilot tones & $25$    \\ \hline
		Zero padding/FFT size $(N)$ & $106/256$    \\ \hline
		CP size $(L)$ & $32$    \\ \hline
		Distance between terminals& { $225$ cm}  \\ \hline
	\end{tabular}
\vspace{-0.2cm}
\end{table}
Overall symbol structure is shown in Fig. 11. We insert two identical training sequences  with $128$ samples for both TO and CFO estimation at the beginning of the symbol. In the OFDM frame, the initial $53$, the center and the final $52$ subcarriers are set to $0$. A total $150$ subcarriers are assigned to data and pilot tones. Since the ratio of data subcarriers to pilot tones is $4$, we have a total number of $125$ data and $25$ pilot subcarriers. After the IFFT operation, a CP with $32$ samples is inserted in the time domain. Experiment parameters are summarized in Table V. { The bandwidth, subcarrier spacing and symbol duration are $500$ kHz, $3.333$ kHz and $0.3$ ms, respectively.}
\vspace{-0.4cm}
\subsection{Receiver Details}
{In the presence of CFO $(\varepsilon)$ and TO $(\gamma)$, the received baseband signal in time domain can be given as:
\begin{align}
	{y}_{T}(\tau)
	=\frac{1}{N}\sum_{\beta=0}^{N-1}s_F(\beta) c_F(\beta) e^{j2\pi(\beta+\varepsilon)(\tau+\gamma)/N} + w_T(\tau),
	\label{Eq23}
\end{align}
where $w_T(\tau)$, $\tau=1,2,\cdots,N$ represent the noise samples in the time domain, whose vector form is given as $\mathbf{w}_T=\mathrm{IFFT}\{\mathbf{w}_F\}$.}
\subsubsection{CFO and TO Estimation}
After receiving signals, by using consecutive and identical training sequences, timing offset (TO) and carrier frequency offset (CFO) are estimated in time and frequency domain, respectively. By using sliding windows, which have the same lengths as the training sequence, the similarity between two blocks is obtained by autocorrelation \cite{MIMOBOOK}. {Hence, TO can be ML estimated as:
\begin{equation}
	\hat{\gamma}=\underset{\gamma}{\arg\: \max}{\Bigg\{\frac{\Big|\sum_{i=\gamma}^{N/2-1+\gamma} y_T(\tau+i)y_T^{*}(\tau+N/2+i)\Big|^2}{\Big|\sum_{i=\gamma}^{N/2-1+\gamma} y_T(\tau+N/2+i)\Big|^2}\Bigg\}}.
	\label{Eq24}
\end{equation}}
 After obtaining the start of the frame (TO), by taking FFT of training sequences, CFO is estimated in the frequency domain with the technique proposed by Moose  \cite{MOOSE} {as follows: 
\begin{equation}
{ 
	\hat{\varepsilon}=\frac{1}{2\pi}\mathrm{tan}^{-1}\Bigg\{\frac{\sum_{\beta=0}^{N-1}\Im\{T_1^*(\beta) T_2(\beta)\} }{ \sum_{\beta=0}^{N-1}\Re\{T_1^*(\beta) T_2(\beta)\} }\Bigg\}},
\label{Eq25}
\end{equation}
where $T_l, l=1,2$ is the $l$th received training sequence in the frequency domain.} Finally, the CP is removed and the CFO is compensated.
\subsubsection{Channel Estimation}
After performing FFT, data and pilot tones are separated. By employing pilot tones, channel coefficients are estimated with one-dimensional linear interpolation as in \cite{PRACT}. Finally, the estimated channel coefficients are fed into the detector.
\subsubsection{SNR Estimation}
For SNR estimation, the method proposed in \cite{arslan2003noise} is considered. The estimated SNR values for all target systems can be given as
\begin{equation}
	{\mathrm{SNR}}=\frac{\sum_{\beta=1}^{N} |y_F(\beta)|^2}{\sum_{\beta=1}^{N} |y_F(\beta)-s_F(\beta) h_F(\beta) |^2}.
	\label{Eq26}
\end{equation}

\begin{figure}[!t]
	\centering
	\includegraphics[width=0.8\linewidth ]{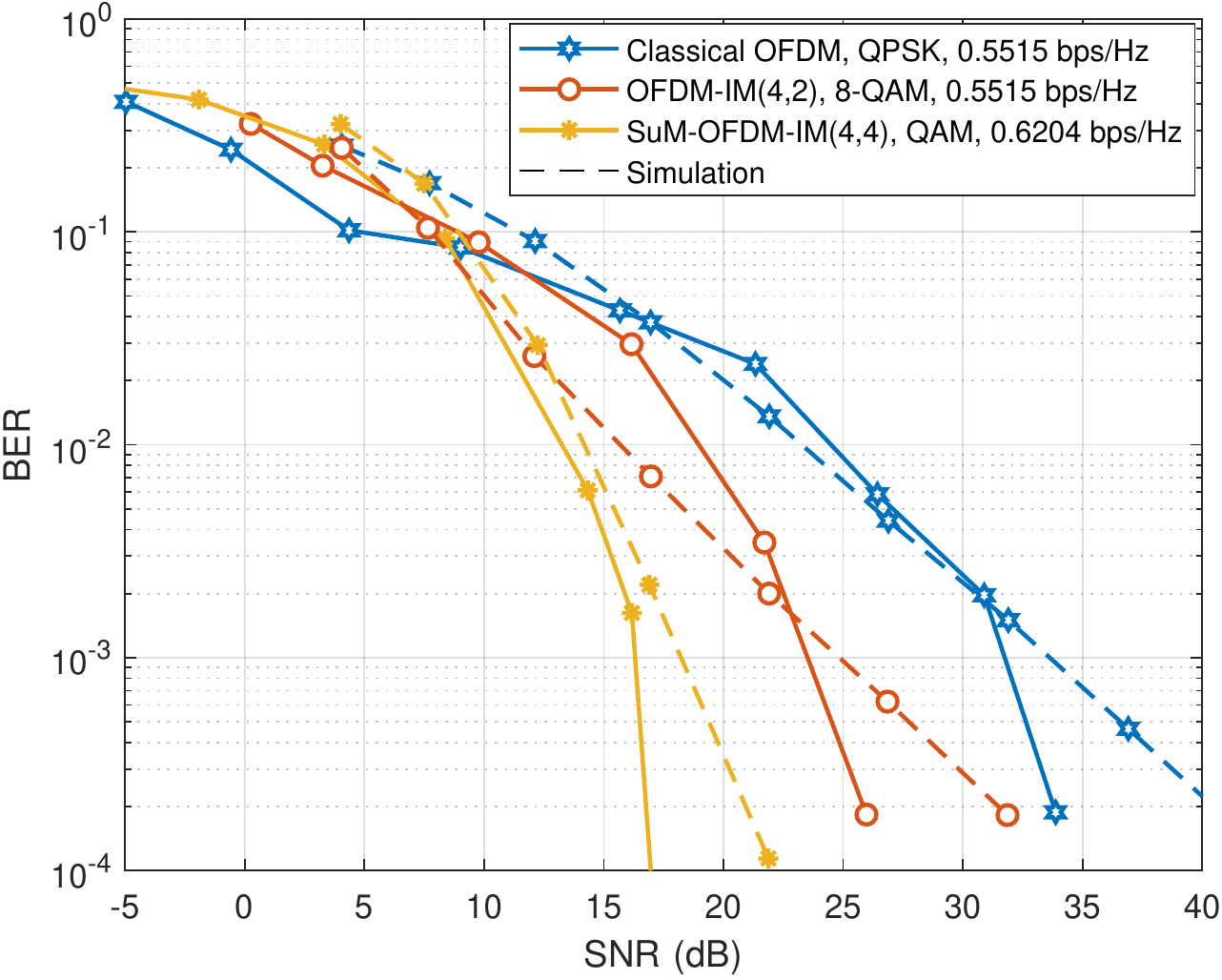}	
	\vspace{-0.2cm}
	\caption{Error performance of OFDM, OFDM-IM and SuM-OFDM-IM under a practical setup {and with computer simulation}.}
	\label{Fig. 10}
	\vspace{-0.5 cm}
\end{figure}
\subsection{Error Performance}
In this subsection, we compare the error performance of SuM-OFDM-IM with OFDM and OFDM-IM under a practical setup. ML detector is employed for all schemes. {We consider a three-tap Rician channel with uniform power delay profile and $K$-factor of $10$ dB to compare simulation results with experimental results similar to \cite{PRACT}}. As seen from Fig. 12, SuM-OFDM-IM outperforms OFDM and OFDM-IM in terms of error performance even though providing a $12.5\%$ higher spectral efficiency. These experimental results are also consistent with the simulation results.  

\vspace{-0.2cm}
\section{Conclusion}	
In this paper, we have proposed SuM-OFDM-IM, which jointly determines MAPs and SAPs by IM to convey additional information bits and replicates the conventional QAM/PSK symbols to obtain a diversity gain. {To enrich the contribution, the separate selection of MAPs and SAPs have also been investigated.} A reduced complexity ML detector, which achieves the same error performance as the ML detector, has been designed and a complexity analysis has been given. Finally, the error performance of the proposed scheme has been investigated both with computer simulations and SDRs under practical impairments. Our exhaustive simulation and practical results demonstrate that the proposed scheme remarkably outperforms OFDM and other OFDM-IM based reference systems in terms of error performance while providing a high spectral efficiency. Consequently, SuM-OFDM-IM can be a possible waveform candidate for 5G and beyond eMBB services due to its appealing advantages such as high spectral efficiency, reliability and diversity gain \cite{6G}. Other precoding techniques can be employed to further increase the diversity order of SuM-OFDM-IM. Moreover, the spectral efficiency can be further enhanced by selecting more than two modes; however, it is left as a future study.
\balance
\bibliographystyle{IEEEtran}
\bibliography{bib}
\begin{IEEEbiography}[{\includegraphics[width=1in,height=1.25in,clip,keepaspectratio]{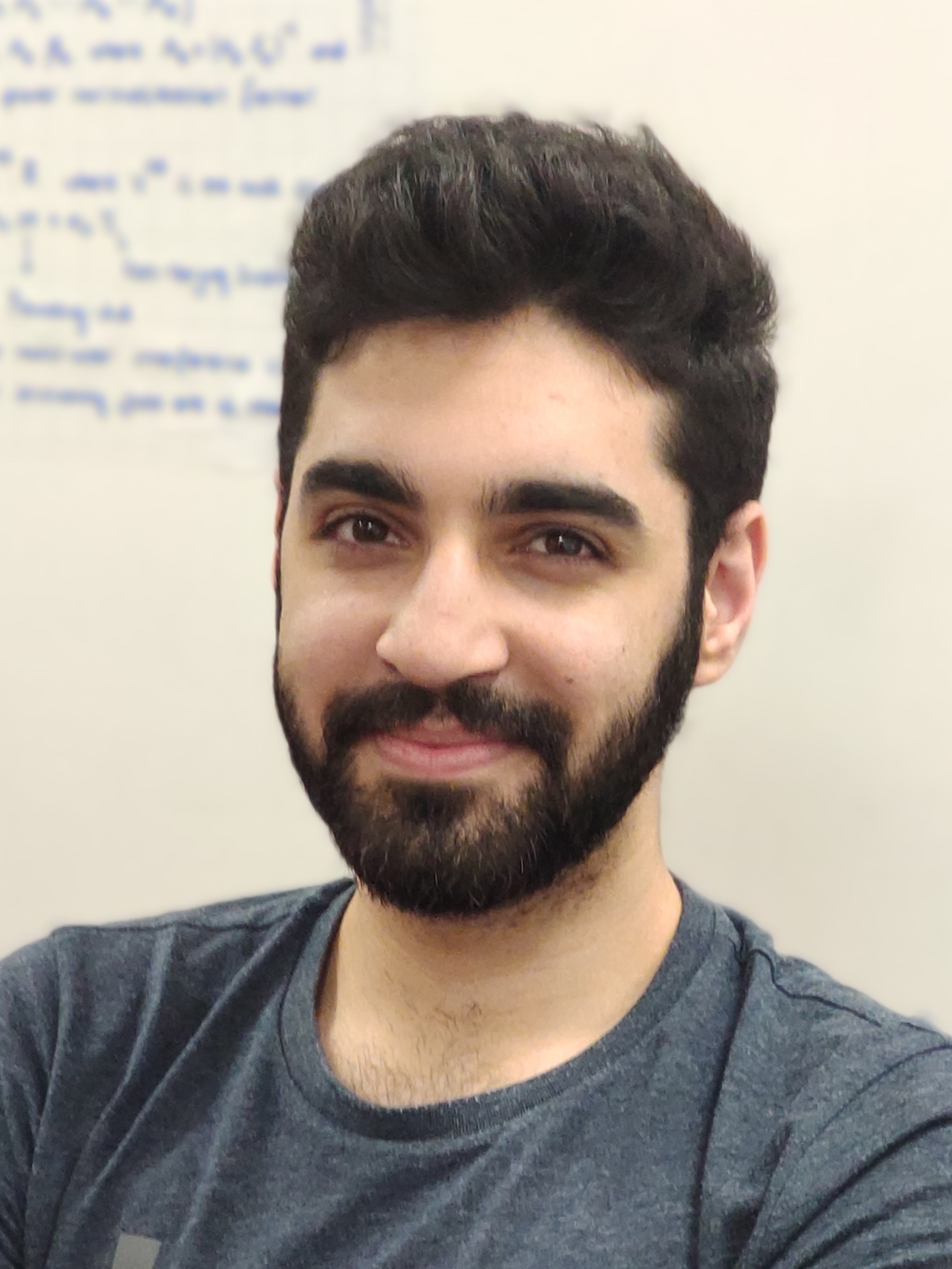}}]{Ali Tugberk Dogukan}
	(S'20) received the B.S. degree from Istanbul Technical University, Istanbul, Turkey, in 2018, and the M.S. degree from Koç University, Istanbul, Turkey, in 2020. He is currently a Research and Teaching Assistant at Koç University while pursuing his Ph.D. degree at the same university. His research interests include waveform design, signal processing for communications, index modulation, and software defined radio-based practical implementation. He was a Reviewer for IEEE.
\end{IEEEbiography}
\begin{IEEEbiography}[{\includegraphics[width=1in,height=1.25in,clip,keepaspectratio]{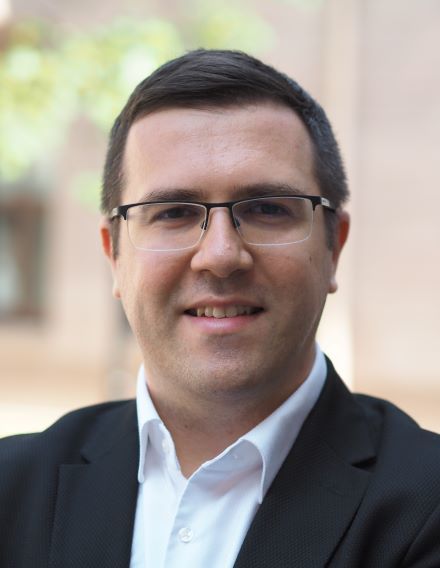}}]{Ertugrul Basar}
	 (S'09-M'13-SM'16) received the B.S. degree (Hons.) from Istanbul University, Turkey, in 2007, and the M.S. and Ph.D. degrees from Istanbul Technical University, Turkey, in 2009 and 2013, respectively. He is currently an Associate Professor with the Department of Electrical and Electronics Engineering, Koç University, Istanbul, Turkey and the director of Communications Research and Innovation Laboratory (CoreLab). His primary research interests include MIMO systems, index modulation, intelligent surfaces, waveform design, visible light communications, and signal processing for communications. 
	
	Recent recognition of his research includes the IEEE Communications Society Best Young Researcher Award for the Europe, Middle East, and Africa Region in 2020, Science Academy (Turkey) Young Scientists (BAGEP) Award in 2018, Turkish Academy of Sciences Outstanding Young Scientist (TUBA-GEBIP) Award in 2017, and the first-ever IEEE Turkey Research Encouragement Award in 2017. 
	
	Dr. Basar currently serves as a Senior Editor of the IEEE Communications Letters and the Editor of the IEEE Transactions on Communications, Physical Communication (Elsevier), and Frontiers in Communications and Networks.
\end{IEEEbiography}

\end{document}